\documentclass[pre,twocolumn,amsmath,amssymb,nofootinbib,floatfix,superscriptaddress]{revtex4}

\usepackage{graphicx,bm,float, physics}
\usepackage{xcolor}
\usepackage[normalem]{ulem}

\makeatletter
\def\graphicscale{\twocolumn@sw{0.3}{0.4}}
\def\graphicthreescale{\twocolumn@sw{0.3}{0.4}}

\begin{document}

\title{Phase diagram and Higgs phases 
of 3D lattice SU($N_c$) gauge
  theories \\with multiparameter scalar potentials}

\author{Claudio Bonati} 
\affiliation{Dipartimento di Fisica dell'Universit\`a di Pisa 
       and INFN, Pisa, Italy}

\author{Alessio Franchi} 
\affiliation{Dipartimento di Fisica dell'Universit\`a di Pisa 
       and INFN, Pisa, Italy}
       
\author{Andrea Pelissetto}
\affiliation{Dipartimento di Fisica dell'Universit\`a di Roma Sapienza
        and INFN, Roma, Italy}

\author{Ettore Vicari} 
\affiliation{Dipartimento di Fisica dell'Universit\`a di Pisa
       and INFN, Pisa, Italy}

\date{\today}

\begin{abstract}
We consider three-dimensional lattice SU($N_c$) gauge theories with
degenerate multicomponent ($N_f>1$) complex scalar fields that
transform under the fundamental representation of the gauge SU($N_c$)
group and of the global U($N_f$) invariance group, interacting with
the most general quartic potential compatible with the global (flavor)
and gauge (color) symmetries.  We investigate the phase diagrams,
identifying the low-temperature Higgs phases and their global and
gauge symmetries, and the critical behaviors along the different
transition lines.  In particular, we address the role of the quartic
scalar potential, which determines the Higgs phases and the
corresponding symmetry-breaking patterns.  Our study is based on the
analysis of the minimum-energy configurations and on numerical Monte
Carlo simulations. Moreover, we investigate whether some of the
transitions observed in the lattice model can be related to the
behavior of the renormalization-group flow of the continuum field
theory with the same symmetries and field content around its stable
{\em charged} fixed points. For $N_c = 2$, numerical results are
consistent with the existence of {\em charged} critical behaviors for
$N_f > N_f^\star$, with $20 < N_f^\star < 40$.
\end{abstract}

\maketitle


\section{Introduction}
\label{intro}

Gauge symmetries provide a unifying theme of contemporary theoretical
physics, describing the dynamics of the Standard Model of fundamental
interactions~\cite{Wilson-74,ZJ-book,Weinberg-book} and critical
phenomena in condensed matter
physics~\cite{Wegner-71,Anderson-book,Sachdev-19}. The interplay
between local gauge and global symmetries is a crucial determinant of
the different phases occurring in gauge
models~\cite{GG-72,OS-78,FS-79,DRS-80,BN-87} and of the thermal and
quantum transitions between them~\cite{BPV-19,BPV-20-su,BPV-20-on}.

We address these issues in three-dimensional (3D) lattice scalar gauge
theories with SU($N_c$) gauge invariance (we name it color gauge
symmetry) and U($N_f$) global (flavor) invariance.  The multicomponent
scalar fields ($N_f> 1$) transform under the fundamental
representation of both groups.  We extend earlier
results~\cite{BPV-19,BPV-20-su}, considering the most general quartic
scalar potential compatible with the SU($N_c$) gauge symmetry and the
global U($N_f$) flavor symmetry. In this extended model, different
low-temperature Higgs phases emerge when varying the potential
parameters. We determine the phase diagrams, focusing, in particular,
on the nature of the low-temperature Higgs
phases~\cite{GG-72,FS-79,BN-87}, and of the phase transitions that
separate the different phases, which are related to the spontaneous
breaking of the global symmetry. All these properties depend on the
parameters of the quartic scalar potential and on the numbers of
colors and flavors, $N_c$ and $N_f$, respectively. In particular, the
phase diagrams for $N_c=2$ and $N_c>2$ are qualitatively different,
because of the presence of an enlarged global symmetry for $N_c=2$ in
the absence of the scalar potential \cite{BPV-19,BPV-20-su}. Moreover,
the phase behavior differs for $N_f<N_c$, $N_f=N_c$, and $N_f>N_c$. In
particular, distinct low-temperature Higgs phases, associated with
different gauge-symmetry breaking patterns, exist only when $N_f\ge
N_c$. We mention that similar studies have been reported for SU($N_c$)
scalar gauge models in which the scalar fields transform under the
adjoint representation of the gauge
group~\cite{SSST-19,SPSS-20,BFPV-21-3d}.

We only consider the case $N_f>1$. The 3D SU(2) gauge theory coupled
to a single scalar SU(2) doublet ($N_f=1$ in our notation) has been
much investigated, due to its relevance for the finite-temperature
electroweak phase transition~\cite{Nadkarni:1989na, Kajantie:1993ag,
  Buchmuller:1994qy, Kajantie:1996mn, Hart:1996ac}. We only mention
that the phase diagram of the single-flavor model shows only a single
phase~\cite{OS-78, FS-79, DRS-80}, indeed the high- and
low-temperature regions turn out to be analytically connected.

An important issue in the present context is the relation between the
statistical gauge model and the corresponding field theory, i.e., the
field theory with the same field content and the same gauge and global
symmetries. In particular, one would like to identify the continuous
transitions that can be described by a {\em charged} fixed point (FP)
of the renormalization-group (RG) flow of the continuum SU($N_c$)
gauge field theory, i.e., with a nonzero gauge-coupling value at the
FP.  At present, for 3D scalar models, this identification has been
done only for Abelian gauge theories \cite{BPV-21-nc,BPV-20-hc}.  No
analogous result has been yet reported in the literature for
non-Abelian gauge models.  In this work we address this issue in the
context of scalar SU($N_c$) gauge models.  Close to four dimensions,
stable charged FPs exist for any $N_c$ and sufficiently large
$N_f$. We numerically investigate this issue for $N_c = 2$.
Finite-size scaling (FSS) analyses of Monte Carlo (MC) simulations
allow us to identify continuous transitions for $N_f = 40$.  They
become of first order in the infinite-gauge coupling limit, in which
gauge fields can be integrated out. This suggests that SU(2) gauge
fields play a role at the transition and thus it seems natural to
associate them with the charged field-theory FP. Since no continuous
transition is found for $N_f=20$, our results suggest that 3D charged
critical behaviors for $N_c = 2$ develop for $N_f > N^\star_f$, with
$20< N^\star_f< 40$.

The paper is organized as follows. In Sec.~\ref{model} we define the
lattice SU($N_c$) gauge model with $N_f$ scalar fields in the
fundamental representation. In Sec.~\ref{obsfss} we introduce the
observables and discuss their FSS behavior, which will be at the basis
of our numerical analyses. In Sec.~\ref{gausym} we discuss the
structure of the Higgs phases that emerge from an analysis of the
mininum-potential configurations, and characterize their global and
gauge symmetry-breaking patterns.  In Sec.~\ref{sft} we discuss the RG
flow of the continuum SU($N_c$) gauge theories with a multiflavor
scalar field and U($N_f$) global symmetry, identifying a stable
charged FP for large $N_f$ at any fixed $N_c$ close to four
dimensions.  In Sec.~\ref{phadia} we discuss the possible phase
diagrams in the space of the Hamiltonian parameters and of the
temperature, for the three cases $N_f<N_c$, $N_f=N_c$, and $N_f>N_c$
emerged in Sec.~\ref{gausym}.  In Sec.~\ref{numresnc2} we present a
numerical study for $N_c=2$ and $N_f=2,20,40$. We perform a FSS
analysis of MC data, to verify the theoretical predictions.  Finally,
in Sec.~\ref{conclu} we summarize and draw our conclusions. A short
discussion of the model for infinite gauge coupling is given in
App.~\ref{AppA}.  Some details on the MC simulations and numerical
analyses are reported in App.~\ref{mcsim}.

\section{Lattice SU($N_c$) gauge models with multiflavor scalar fields}
\label{model}

We consider lattice scalar gauge models with SU($N_c$) local
invariance defined on cubic lattices of linear size $L$ with periodic
boundary conditions. The fundamental fields are complex matrices
$\Phi^{af}_{\bm x}$, with $a=1,...,N_c$ ({\em color} index) and
$f=1,...,N_f$ ({\em flavor} index), defined on the lattice sites and
${\rm SU}(N_c)$ matrices $U_{{\bm x},\mu}$ defined on the lattice
links.  The partition function is
\begin{eqnarray}
  &&  Z = \sum_{\{\Phi,U\}} e^{-\beta H}\,,\qquad \beta=1/T\,,
  \label{partfunc}\\ 
&&  H  = H_K(\Phi,U) + H_V(\Phi) + H_G(U)\,,
\label{hgauge}
\end{eqnarray}
where $H$ is the sum of the scalar-field kinetic term $H_K$, of the
local scalar potential $H_V$, and of the pure-gauge Hamiltonian $H_G$.
As usual, we set the lattice spacing equal to one, so that all lengths
are measured in units of the lattice spacing.  The kinetic term $H_K$
is given by
\begin{eqnarray}
  H_K(\Phi,U) =   - J
  N_f \sum_{{\bm x},\mu} {\rm Re} \, {\rm Tr} \,\Phi_{\bm x}^\dagger \,
 U_{{\bm x},\mu} \, \Phi_{{\bm x}+\hat{\mu}}^{\phantom\dagger}\,.
  \label{Kinterm}
\end{eqnarray}
In the following we set  $J=1$, so that energies are
measured in units of $J$. The second term $H_V$ is 
\begin{eqnarray}
  &&H_V(\Phi) = \sum_{\bm x} V(\Phi_{\bm
    x})\,,\label{potential}\\ &&V(\Phi)= {r\over 2} \, {\rm
    Tr}\,\Phi^\dagger\Phi + {u\over 4} \left( {\rm
    Tr}\,\Phi^\dagger\Phi\right)^2 + {v\over 4} \, {\rm
    Tr}\,(\Phi^\dagger\Phi)^2 \,.\nonumber
\end{eqnarray}  
The potential $V(\Phi)$ is the most general quartic polynomial that is
symmetric under [U($N_f$)$\otimes$U($N_c$)]/U(1) transformations.  For
$v=0$ the symmetry of the scalar potential enlarges to O($M$) with
$M=2N_f\,N_c$.  Finally, we define \cite{Wilson-74}
\begin{eqnarray}
&&H_G(U) =
  - {\gamma\over N_c} \sum_{{\bm x},\mu>\nu}  {\rm Re} \,
  {\rm Tr}\, \Pi_{{\bm x},\mu\nu}\,,
\label{plaquette}\\
&&\Pi_{{\bm x},\mu\nu}=
      U_{{\bm x},\mu} \,U_{{\bm x}+\hat{\mu},\nu} \,U_{{\bm
    x}+\hat{\nu},\mu}^\dagger \,U_{{\bm x},\nu}^\dagger \,,
\nonumber
\end{eqnarray}
where the parameter $\gamma$ plays the role of inverse gauge coupling. 

The Hamiltonian $H$ is invariant under local SU($N_c$) and global
U($N_f$) transformations. Under these transformations, the scalar
field transforms under the fundamental representation of both groups.
Note that U($N_f$) is not a simple group and thus we may separately
consider SU($N_f)$ and U(1) transformations, that correspond to
$\Phi^{af} \to \sum_g V^{fg} \Phi^{ag}$, $V\in$ SU($N_f$), and
$\Phi^{af} \to e^{i\alpha} \Phi^{ag}$, $\alpha \in [0,2\pi)$,
  respectively.  Note that, since the diagonal matrix with entries
  $e^{i 2\pi/N_c}$ is an SU($N_c$) matrix, $\alpha$ can be restricted
  to $[0,2\pi/N_c)$ and the global symmetry group is more precisely
    U$(N_f)/\mathbb{Z}_{N_c}$.

In this study we consider fixed-length fields satisfying
\begin{equation}
{\rm Tr}\, \Phi_{\bm x}^\dagger \Phi_{\bm x}^{\phantom\dagger} = 1\, 
\label{trphi2}
\end{equation}
and the lattice Hamiltonian
\begin{eqnarray}
&&H = - N_f\sum_{{\bm x},\mu} {\rm Re}\,{\rm Tr} \,\Phi_{\bm x}^\dagger \,
U_{{\bm x},\mu} \, \Phi_{{\bm x}+\hat{\mu}}^{\phantom\dagger}
\label{hfixedlength}  \\
&&\quad+ {v\over 4} \sum_{\bm x}  {\rm Tr}\,(\Phi_{\bm x}^\dagger\Phi_{\bm x})^2
- {\gamma\over N_c} \sum_{{\bm x},\mu>\nu}  {\rm Re} \, {\rm Tr}\,
\Pi_{{\bm x},\mu\nu}\,.\nonumber
\end{eqnarray}
This model can be formally obtained from the general one by
considering the limit $u\to\infty$ keeping the ratio $r/u=-1$ fixed.
We expect it to have the same features of models with generic values
of $r$ and $u$.

\section{Observables, order parameter and finite-size scaling}
\label{obsfss}

In our simulations we compute the energy density and the specific
heat, defined as
\begin{eqnarray}
\label{ecvdef}
E = -\frac{1}{3 {\cal V}} \langle H \rangle\,,\quad
C_V =\frac{1}{\cal V}\left( \langle H^2 \rangle 
- \langle H  \rangle^2\right)\,,\quad
\end{eqnarray}
where ${\cal V}=L^3$ is the volume of the lattice.

To study the breaking of the global U($N_f$) symmetry, we monitor
correlation functions of the gauge-invariant bilinear operator
\begin{equation}
  A_{\bm x}^{fg} = \sum_a {\bar\Phi}_{\bm x}^{af} \Phi_{\bm x}^{ag}\,,
  \qquad
  Q_{\bm x}^{fg} = A_{\bm x}^{fg} - {1\over N_f} \delta^{fg}\,,
\label{qdef}
\end{equation}
which is invariant under the U(1) global transformations and satisfies
${\rm Tr} \, A_{\bm x}=1$ and ${\rm Tr} \, Q_{\bm x}=0$, because of
the fixed-length constraint ${\rm Tr} \, \Phi^\dagger \Phi=1$.  We
define its two-point correlation function (since we use periodic
boundary conditions, translation invariance holds)
\begin{equation}
G({\bm x}-{\bm y}) = \langle {\rm Tr}\, Q_{\bm x} Q_{\bm y} \rangle\,,
\label{gxyp}
\end{equation}
the corresponding susceptibility $\chi=\sum_{\bm x} G({\bm x})$,
and the second-moment correlation length
\begin{eqnarray}
\xi^2 = {1\over 4 \sin^2 (\pi/L)} {\widetilde{G}({\bm 0}) -
  \widetilde{G}({\bm p}_m)\over \widetilde{G}({\bm p}_m)}\,,
\label{xidefpb}
\end{eqnarray}
where ${\bm p}_m = (2\pi/L,0,0)$ and $\widetilde{G}({\bm
  p})=\sum_{{\bm x}} e^{i{\bm p}\cdot {\bm x}} G({\bm x})$ is the
Fourier transform of $G({\bm x})$.

To monitor the breaking of the U(1) global symmetry,
we consider gauge-invariant operators that transform nontrivially
under these transformations. For $N_c = 2$, 
we consider the bilinear operator
\begin{equation}
  Y_{\bm x}^{fg} = \epsilon^{ab}\, \Phi_{\bm x}^{af}\, \Phi_{\bm x}^{bg} \,,
  \label{Yphase}
\end{equation}
where $\epsilon^{ab}$ is the completely antisymmetric tensor in the
color space, with $\epsilon^{12}=1$.  For $N_f=2$, $Y_{\bm x}^{fg}$ is
equivalent to the determinant of $\Phi_{\bm x}$:
\begin{equation}
  Y_{\bm x}^{fg} = \epsilon^{fg} D_{\bm x}\,, \;\; D_{\bm x}\equiv
  {\rm det}\,\Phi_{\bm x}\,,
  \quad {\rm for}\; N_c=N_f=2\,.
\label{Ydet}
\end{equation}
We define the corresponding two-point correlation function
\begin{equation}
  G_Y({\bm x}-{\bm y}) = \langle \, {\rm Tr}
  \,Y_{\bm x}^\dagger Y_{\bm y} \rangle\,,  
\label{gyxyp}
\end{equation}
the susceptibility $\chi_Y=\sum_{\bm x} G_Y({\bm x})$ and the
second-moment correlation length $\xi_Y$ as in Eq.~(\ref{xidefpb}).
For $N_c=N_f=2$, $G_Y$ can be written as
 \begin{equation}
   G_Y({\bm x}-{\bm y}) = 2 \langle \, \bar{D}_{\bm x} D_{\bm y}
   \rangle\,.  
\label{gyxyp2}
\end{equation}
In our numerical study we also consider the Binder parameter
\begin{equation}
U = \frac{\langle \mu_2^2\rangle}{\langle \mu_2 \rangle^2} \,, \qquad
\mu_2 = \frac{1}{{\cal V}^2}  
\sum_{{\bm x},{\bm y}} {\rm Tr}\,Q_{\bm x} Q_{\bm y}\,,
\label{binderdef}
\end{equation}
and the ratio
\begin{equation}\label{rxidef}
R_{\xi}=\xi/L\,.
\end{equation}
Analogous quantities $U_Y$ and $R_{\xi,Y}$ can be defined using
the correlations of the operator $Y$ defined in Eq.~(\ref{Yphase}).

At a continuous phase transition, any RG invariant ratio $R$, such as
the Binder parameters $U$ and $U_Y$ or the ratios $R_\xi$ and
$R_{\xi,Y}$, scales as~\cite{PV-02}
\begin{eqnarray}
R(\beta,L) = f_R(X) +  L^{-\omega} g_R(X) + \ldots \,, \label{scalbeh}
\end{eqnarray}
where 
\begin{equation}
  X = (\beta-\beta_c)L^{1/\nu}\,.
  \label{Xdef}
  \end{equation}
The function $f_R(X)$ is universal up to a multiplicative rescaling of
its argument, $\nu$ is the correlation-length critical exponent, and
$\omega$ is the exponent associated with the leading irrelevant
operator.  In particular, $R^*\equiv f_R(0)$ is universal, depending
only on the boundary conditions and aspect ratio of the lattice.
Since $R_\xi$ defined in Eq.~\eqref{rxidef} is an increasing function
of $\beta$, we can combine the RG predictions for $U$ and $R_\xi$ to
obtain
\begin{equation}
  U(\beta,L) = F(R_\xi) + O(L^{-\omega})\,,
\label{uvsrxi}
\end{equation}
where $F$ now depends on the universality class, boundary conditions,
and lattice shape, without any nonuniversal multiplicative factor.
Eq.~\eqref{uvsrxi} is particularly convenient because it allows one to
test universality-class predictions without requiring a tuning of
nonuniversal parameters.

The Binder parameter $U$ is also useful to identify weak first-order
transitions, when large lattice sizes are required to observe a finite
latent heat and bimodal energy distributions.  Indeed, while $U$ is
bounded as $L\to \infty$ at a continuous transition, at a first-order
transition its maximum $U_{\rm max}$ increases as the
volume~\cite{CLB-86,VRSB-93,CPPV-04}, i.e.,
\begin{equation}\label{Ufirst}
U_{\rm max}= a {\cal V} + O(1)\,,\qquad {\cal V}=L^3\,.
\end{equation}
Therefore, $U$ has a qualitatively different scaling behavior for
first- and second-order transitions.  The absence of a data collapse
in plots of $U$ versus $R_{\xi}$ may be considered as an early
indication of the first-order nature of the transition~\cite{PV-19}.
We also recall that, according to the standard phenomenological
theory~\cite{CLB-86}, the maximum value $C_{\rm max}(L)$ of the
specific heat at first-order transitions is expected to asymptotically
increases as
\begin{eqnarray}
  C_{\rm max}(L) = {\cal V} \left[
    {1\over 4} \Delta_h^2 + O(1/{\cal V})\right]\,,
  \label{cmaxsc}
\end{eqnarray}  
where $\Delta_h$ is the latent heat, defined as $\Delta_h =
E(\beta\to\beta_c^+) - E(\beta\to\beta_c^-)$. Moreover the $\beta$
values at the maximum of the specific heat converge to the transition
point as $\beta_{{\rm max},C}(L)-\beta_c\approx c\,{\cal V}^{-1}$.

\section{Low-temperature Higgs phases}
\label{gausym}

The lattice gauge models we consider may have different Higgs phases
associated with different symmetry-breaking patterns. They are
determined by the minima of the scalar potential
\begin{equation}
V(\Phi) = {v\over 4} {\rm Tr}\,(\Phi^\dagger \Phi)^2 \,.
\label{locpotential}
\end{equation}
In this section we discuss the main properties of these phases, which
depend on the parameter $v$, the number of colors $N_c$ and of flavors
$N_f$.  It is worth noting that this discussion applies to generic
$D$-dimensional systems, therefore also to $D=4$ space-time systems
that may be relevant in the context of high-energy physics.

\subsection{Configurations in the zero-temperature limit}
\label{zetotconf}

For $\beta\to\infty$, the relevant configurations are those that
minimize $V(\Phi)$. To determine the minima, we use the singular-value
decomposition that allows us to rewrite the field $\Phi$ as
\begin{equation}
  \Phi^{af} = \sum_{bg} C^{ab} W^{bg} F^{gf}\,,
  \label{singdec}
\end{equation}
where $C\in {\rm U}(N_c)$ and $F\in {\rm U}(N_f)$ are unitary
matrices, and $W$ is an $N_c \times N_f$ rectangular matrix. Its
nondiagonal elements vanish ($W^{ij} = 0$ for $i\not= j$), while the
diagonal elements are real and nonnegative, $W^{ii} = w_i>0$
($i=1,...,q$),
\begin{equation}
  q={\rm Min}[N_f,N_c]\,.
  \label{qqdef}
\end{equation}
Substituting the expression (\ref{singdec}) 
in $V(\Phi)$, we obtain
\begin{equation}
   V(\Phi) = \frac{v}{4} \sum_{i=1}^q w_i^4\ .
\end{equation}
A straightforward minimization of this expression, subject to the
constraint
\begin{equation}
\hbox{Tr}\,
\Phi^\dagger \Phi = \sum_{i=1}^q w_i^2 = 1\,, 
\label{constraint}
\end{equation}
gives two different solutions, that depend on the sign of $v$:
\begin{eqnarray}
& \hbox{(I)} \;\; &w_1 = 1\,, \quad w_{2}=...=w_q = 0\,, \quad 
{\rm for}\;v<0\,,\qquad \label{solution1} \\
&\hbox{(II)} \;\;  &w_1 = \ldots = w_q = {1/\sqrt{q}}\,,
\quad {\rm for}\;v>0\,.\quad
\label{solution2}
\end{eqnarray}
Analogous results hold for the general potential (\ref{potential}). 
If we perform 
the substitution (\ref{singdec}), we obtain the 
potential of a $q$-component model with cubic anisotropy
\begin{equation}
V(\Phi)=\frac{r}{2}\left(\sum_i w_i^2\right) + 
\frac{u}{4}\left(\sum_i w_i^2\right)^2 +
\frac{v}{4}\left(\sum_i w_i^4\right)\,.
\end{equation}
The minimum of the potential is $w_1 = \ldots w_q = 0$ for $r > 0$.
It corresponds to the diagonally ordered state $w_1=\ldots =w_q>0$ for
$r<0$ and $v>0$ (and $u+v/q>0$ for stability), and to the axis-aligned
state $w_1>0$, $w_2=\ldots=w_q=0$, for $r < 0$ and $v<0$ (and $u+v>0$
for stability).

For solutions of type (I), we can rewrite the field as 
\begin{equation}
  \Phi^{af} = s^a z^f\,,
\label{minvlt0}
\end{equation}
where $s$ and $z$ are unit-length complex vectors of dimension $N_c$
and $N_f$, respectively, satisfying $\bar{\bm s}\cdot {\bm s}=1$ and
$\bar{\bm z}\cdot {\bm z}=1$.

For solutions of type (II), we have instead
\begin{equation}
  \Phi^{af} = {1\over \sqrt{q}}
  \sum_{k=1}^q C^{ak} F^{kf}.
\label{Phi-vgt0}
\end{equation}
This expression can be further simplified, parameterizing $\Phi$ in
terms of a single unitary matrix.  If $N_f\ge N_c$, thus $q=N_c$, we
can rewrite Eq.~(\ref{Phi-vgt0}) as
\begin{equation}
\Phi^{af} = 
{1\over \sqrt{N_c}}\sum_{g=1}^{N_f} \widehat{C}^{ag} F^{gf} \,,
\label{phicf}
\end{equation}
where $\widehat{C} = C\oplus I_{N_f-N_c}$ is an $N_f$-dimensional
unitary matrix ($I_{p}$ is the $p$-dimensional identity matrix).
Since $\widehat{C}$ is a unitary matrix, we can express $\Phi$ in
terms of a single unitary matrix $F' = \widehat{C} F$, i.e., we can
set $C = I$ in Eq.~(\ref{Phi-vgt0}). Due to gauge invariance, $F$ is
an element of U($N_f$)/SU($N_c$).

If $N_f \le N_c$, thus $q=N_f$, we can repeat the same argument to
show that one can set
\begin{equation}
  F = I\,,\qquad \Phi^{af} = q^{-1/2}\,C^{af}\,,
  \label{FPhi}
\end{equation}
without loss of generality.  Then, we can use the SU($N_c$) gauge
transformations to further simplify the expression of $\Phi^{af}$,
obtaining
\begin{equation}
  \Phi^{af} = {1\over \sqrt{N_f}} \,\phi \,\delta^{af}\,,
\label{minvgt0}
\end{equation}
where $\phi$ is a phase satisfying $|\phi| = 1$.  For $N_f < N_c$, the
phase $\phi$ can be eliminated by performing an appropriate SU($N_c$)
gauge transformation~\cite{BPV-19,BPV-20-su}.  Indeed, let us define
the SU$(N_c)$ matrix $V = \hbox{diag }(g_1,\ldots,g_{N_c})$ with $g_a
= \phi$ for $1 \le a \le N_f$, $g_a = \phi^{-N_f}$ for $a =N_f+1$, and
$g_a = 1$ for $a > N_f + 1$.  Then, we have
\begin{equation}
  \Phi^{af} = {1\over \sqrt{N_f}} \,\phi \,\delta^{af} = 
   {1\over \sqrt{N_f}} \sum_{ab} V^{ab} \delta^{bf}.
\end{equation}
Therefore, for $N_f<N_c$ a 
representative of the mininum configurations is simply
\begin{equation}
\Phi^{af} = {1\over \sqrt{N_f}}\,\delta^{af}\,.
\label{minvgt0nfltnc}
\end{equation}

To distinguish the nature of the zero-temperature configurations, one
can use the bilinear operator $A_{\bm x}$ defined in
Eq.~(\ref{qdef}). If the field is parametrized as in
Eq.~(\ref{singdec}), we have
\begin{equation}
\hbox{Tr}\, A^2 = \sum_{i=1}^q w_i^4,
\end{equation}
so that 
\begin{eqnarray}
\hbox{(I)}\;\;  \hbox{Tr}\, A^2 = 1, \qquad 
\hbox{(II)}\;\; \hbox{Tr}\, A^2 = {1\over q}\,,
\label{B2-predictions}
\end{eqnarray}
for solutions of type (I) and (II), respectively.

We now discuss the large-$\beta$ behavior of the gauge fields.  If we
minimize the kinetic term (\ref{Kinterm}), we obtain
\begin{equation}
\Phi_{\bm x} = U_{{\bm x},\mu} \Phi_{{\bm x}+\hat{\mu}}\,.
\end{equation}
Repeated applications of this relation along a plaquette lead to the
equation $\Phi_{\bm x} = \Pi_{\bm x} \Phi_{\bm x}$.  For minimum
configurations of type (I), using Eq.~(\ref{minvlt0}), we have
\begin{equation}
s_{\bm x}^{a} = \sum_b \Pi_{\bm x}^{ab} s_{\bm x}^b \,,
\label{minplaq}
\end{equation}
i.e., $\Pi_{\bm x}$ has necessarily a unit eigenvalue.  Thus, for
$\beta\to\infty$ there is still a residual dynamics of the gauge
fields, leading to a pure SU($N_c-1$) gauge model with Hamiltonian
$H_G(U)$. If the relevant configurations are those of type (II), see
Eq.~(\ref{solution2}), $\Pi_{\bm x}$ has $q$ unit eigenvalues, which
further reduce the dynamics of the gauge fields.  In particular, for
$N_f \ge N_c$, $\Pi_{\bm x} = 1$ and the gauge variables are gauge
equivalent to the trivial configuration, i.e. $U_{{\bm x},\mu} =
V_{\bm x}^\dagger V_{{\bm x}+\hat{\mu}}$ where $V_{\bm x}\in$
SU($N_c$). This is true in a finite volume too, since the same
argument can be used to prove that also Polyakov loops winding around
the lattice converge to the identity as $\beta\to\infty$.

In our discussion we have assumed that the relevant scalar-field
configurations in the large-$\beta$ limit are only determined by the
potential term $S_V(\Phi)$, as long as $v\neq 0$.  We show in
App.~\ref{AppA} that this occurs for $\gamma = 0$ and $N_c=2$, but we
expect this to be a general result, as in the case of the analogous
model in which the scalar fields transform in the adjoint
representation of the gauge group (see the appendix of
Ref.~\cite{BFPV-21-3d}).  For $v=0$ the minimum configurations are
determined by the minima of the kinetic term $S_K(\Phi,U)$. For
$N_c\ge 3$ numerical results~\cite{BPV-19,BPV-20-su} show that the
relevant configurations correspond to solution (I), so that the fields
can be parametrized as in Eq.~(\ref{minvlt0}). This implies that the
behavior is the same as for $v < 0$. For $N_c = 2$, the large-$\beta$
behavior for $v=0$ differs from that for $v\ne 0$, because of the
global symmetry enlargement, as discussed in the Appendix of
Ref.~\cite{BPV-20-su} and in the Appendix \ref{AppA} of this work.

\subsection{The model for $v<0$}
\label{negvcase}

For $v<0$ the relevant minimum configurations take the form
(\ref{minvlt0}).  Modulo gauge transformations, they are invariant
under $\hbox{U}(1) \oplus \hbox{U}(N_f-1)$ transformations, leading to
the global-symmetry breaking pattern
\begin{equation}
\hbox{U}(N_f) \to \hbox{U}(1) \oplus \hbox{U}(N_f-1).
\label{gsbp-v-less-0}
\end{equation}
We can also determine the gauge-symmetry breaking pattern, i.e., the
residual gauge symmetry of the minimum-potential configurations, once
$\Phi^{af}$ has been fixed---as the gauge symmetry cannot be
spontaneously broken, this is only possible by adding a suitable gauge
fixing. We obtain
\begin{equation}
{\rm SU}(N_c) \to {\rm  SU}(N_c-1)\,,
\label{gspnegv}
\end{equation}
independently of the flavor number $N_f$.

The symmetry-breaking pattern (\ref{gsbp-v-less-0}) is the same as in
the CP$^{N_f-1}$ model. Thus, if the gauge dynamics is not relevant at
the transition, for $v < 0$ we expect the non-Abelian gauge model with
U($N_f$) global symmetry and the CP$^{N_f-1}$ model to have the same
critical behavior, for any $N_c$.  The correspondence between the two
models can also be established by noting that the relevant order
parameter at the transition is the bilinear combination $A_{\bm x}$
defined in Eq.~(\ref{qdef}).  For minimum configurations, it takes the
form
\begin{equation}
A_{\bm x}^{fg} = \bar{z}_{\bm x}^f z_{\bm x}^g\,, 
\label{bproj}
\end{equation}
i.e., it represents a local projector onto a one-dimensional space.
If we assume that the critical behavior of the gauge model is only
determined by the fluctuations of the order parameter $A_{\bm x}$ that
preserve the minimum-energy structure (\ref{bproj}), the effective
scalar model that describes the critical fluctuations can be
identified with the CP$^{N_f-1}$ model. Indeed, the standard
nearest-neighbor CP$^{N-1}$ action is the simplest action for a local
projector $P^{\alpha\beta}_{\bm x}$:
\begin{eqnarray}
  H_{\rm CP} = - J \sum_{{\bm x},\mu} \hbox{Tr}\, P_{\bm x} P_{{\bm
      x}+\hat{\mu}}\,, \qquad P_{\bm x}^{\alpha\beta} =
  \bar\varphi_{\bm x}^\alpha \varphi_{\bm x}^\beta\,,
\label{srp}
\end{eqnarray}
where $\varphi^\alpha_{\bm x}$ is a unit complex vector.  We recall
that only for $N=2$ does the 3D CP$^{N-1}$ model (\ref{srp}) undergo a
continuous transition, which belongs to the O(3) universality
class. For $N \ge 3$, the model undergoes first-order transitions
\cite{PV-19,PV-19-AH3d,PV-20-ln}, in agreement with a general
Landau-Ginzburg-Wilson (LGW) argument~\cite{PV-19}.  Note, however,
that in some models that are expected to have the same critical
behavior as the CP$^{N-1}$ model and that undergo transitions with the
same symmetry breaking pattern, numerical studies favor a continuous
transition also for $N=3$, see, e.g., Refs.~\cite{KS-12, NCSOS-11,
  NCSOS-13}.  The LGW argument assumes that gauge fields do not play a
role at the transition. If instead gauge fields become critical,
continuous transitions with symmetry breaking pattern
(\ref{gsbp-v-less-0}) are possible. These are controlled by the
charged FP of the Abelian-Higgs field theory
\cite{MZ-03,IZMHS-19}. This occurs for $N>N^\star$ with $N^\star=7(2)$
in the 3D lattice Abelian-Higgs model with noncompact gauge fields
\cite{BPV-21-nc}.

As we have already discussed, since U($N_f$) is not simple, we can
separately break the SU($N_f$) and U(1) subgroups. For $v < 0$,
Eq.~(\ref{gsbp-v-less-0}) implies that we can only observe the
breaking of the SU($N_f$) group. The U(1) subgroup is unbroken in the
whole low-temperature phase.

\subsection{The model for $v>0$}
\label{posvcase}

The critical behavior is more complex for $v>0$, as we must
distinguish three different cases: $N_f<N_c$, $N_f=N_c$, and
$N_f>N_c$.  For $N_f\le N_c$, the minimum-potential configurations
take the form
\begin{eqnarray}
   &&\Phi^{af}= {1\over \sqrt{q}}\,\delta^{af}\,,
    \quad {\rm for}\;N_f<N_c\,,
   \label{phiagnfltltnc}\\
  &&\Phi^{af}= {1\over \sqrt{q}}\,\delta^{af}\phi\,,\quad
  \phi \in {\rm U}(1)\,,  \quad {\rm for}\;N_f=N_c\,.
\nonumber 
\end{eqnarray}
In these cases, we do not expect to observe transitions controlled by
the bilinear operator $Q$ defined in Eq.~(\ref{qdef}).  Indeed, $Q$
vanishes trivially for the configurations given in
Eq.~(\ref{phiagnfltltnc}).  A stronger argument is provided by the
analysis of the global-symmetry breaking pattern.  The global
invariance group of the ordered phase is given by the transformations
$B\in\hbox{U}(N_f)$ such that
\begin{equation}
\sum_g B^{fg} \Phi^{ag} = \sum_b V^{ab} \Phi^{bf}\,, 
\label{eq46}
\end{equation}
for some SU($N_c$) matrix $V$. For $N_f = N_c$, using
Eq.~(\ref{phiagnfltltnc}), we obtain $B = V$, i.e., the global
invariance group is the SU($N_f$) subgroup.  Therefore, for $N_f =
N_c$, the global symmetry-breaking pattern is
\begin{equation}
\hbox{U}(N_f) \to  \hbox{SU}(N_f)\,.
\end{equation}
Thus, transitions associated with the breaking of the U(1) invariance
are possible.

For $N_f < N_c$, $B$ can be any unitary matrix. Indeed, if we take $V
= B \oplus V_2$, where $V_2$ is any unitary matrix of dimension $N_c -
N_f$, such that the product of the determinants of $B$ and $V_2$ is 1,
Eq.~(\ref{eq46}) is satisifed.  Therefore, for $N_f < N_c$ any
U($N_f$) transformation leaves the minimum-potential configurations
invariant.  Thus, there is no global symmetry breaking, and therefore
no transition is expected.

When $N_f>N_c$, 
the minimum-potential configurations take the form
\begin{equation}
  \Phi^{af} = {1\over \sqrt{N_c}}\,F^{af}\,, \qquad F\in {\rm U}(N_f)\, .
  \label{minnfltnc}
  \end{equation}
Moreover, see the discussion following Eq.~(\ref{minplaq}), gauge
configurations are trivial. As before, we assume that in the ordered
phase the relevant fluctuations are those that preserve this
structure.  Therefore, the field $\Phi^{af}_{\bm x}$ can be
parameterized as in Eq.~(\ref{minnfltnc}), with a site-dependent
unitary matrix $F_{\bm x}$, and we can set $U_{{\bm x},\mu} = V_{\bm
  x}^\dagger V_{{\bm x}+\hat{\mu}}$ with $V_{\bm x}\in {\rm SU}(N_c)$.
Substituting this parameterization in the kinetic term of the
Hamiltonian we obtain
\begin{equation}
H_K = - {N_f\over N_c} \sum_{{\bm x}\mu} {\rm Re}\,\hbox{Tr}\, 
  (F^\dagger_{\bm x}  \widehat{V}_{\bm x}^{\dag} Y 
   \widehat{V}_{{\bm x}+\hat{\mu}} F_{{\bm x}+\hat{\mu}} )\,, 
\end{equation}
where $Y = I_{N_c} \oplus 0$ is an $N_f\times N_f$ diagonal matrix in
which the first $N_c$ elements are 1 and the other $N_f-N_c$ elements
are 0, and $\widehat{V} = V \oplus I_{N_f-N_c}$.  This action is
invariant under the global transformations $F_{\bm x}\to F_{\bm x} M$,
with $M\in \mathrm{U}(N_f)$, and under the local transformations
\begin{equation}
\begin{aligned}
& F_{\bm x}\to W_{\bm x}F_{\bm x}\,,\quad 
\widehat{V}_{\bm x}\to \widehat{V}_{\bm x} G_{\bm x}\,,\\
& W_{\bm x}=W^{(1)}_{\bm x}\oplus W^{(2)}_{\bm x}\,,
\quad G_{\bm x}=W^{(1)}_{\bm x}\oplus I_{N_f-N_c}
\,, 
\end{aligned}
\end{equation}
where $W^{(1)}_{\bm x} \in\mathrm{SU}(N_c)$, $W^{(2)}_{\bm x}
\in\mathrm{U}(N_f-N_c)$ ($F_{\bm x}$ is unitary so that $F^{\dag}_{\bm
  x}F_{\bm x}=I_{N_f}$).  The global symmetry of the effective model
that describes the critical fluctuations is therefore
\begin{equation}
\frac{\mathrm{SU}(N_f)}{\mathrm{SU}(N_c)\otimes\mathrm{SU}(N_f-N_c)}\ ,
\end{equation}
which corresponds to the global symmetry-breaking pattern
\begin{equation}
{\rm U}(N_f)\rightarrow \mathrm{SU}(N_c)\otimes\mathrm{U}(N_f-N_c)\ .
\label{glsymbrpat}
\end{equation}

\section{RG flow of the gauge field theory}
\label{sft}

Previous studies of the critical behavior (or continuum limit) of 3D
lattice gauge theories with scalar matter have shown the emergence of
two different scenarios. In some models there are transitions where
scalar-matter and gauge-field correlations are both critical. In this
case the critical behavior is controlled by a charged FP in the RG
flow of the corresponding continuum gauge field
theory~\cite{ZJ-book}. This occurs, for instance, in the 3D lattice
Abelian-Higgs model with noncompact gauge fields \cite{BPV-21-nc}, and
in the compact model with $q$-charged ($q\ge 2$) scalar
fields~\cite{BPV-20-hc}, for a sufficiently large number of
components. Indeed, the critical behavior along one of the transition
lines occurring in these models is associated with the stable FP of
the multicomponent scalar electrodynamics or Abelian-Higgs field
theory~\cite{HLM-74,DHMNP-81,FH-96,YKK-96,MZ-03,KS-08,IZMHS-19},
characterized by a nonvanishing gauge coupling.

Alternatively, it is possible that only scalar-matter correlations are
critical at the transition. The gauge variables do not display
long-range correlations, although their presence is crucial to
identify the gauge-invariant scalar-matter critical degrees of
freedom.  At these transitions, gauge fields prevent non-gauge
invariant scalar correlators from acquiring nonvanishing vacuum
expectation values and developing long-range order: the gauge symmetry
hinders some scalar degrees of freedom---those that are not gauge
invariant---from becoming critical.  In this case the critical
behavior or continuum limit is driven by the condensation of
gauge-invariant scalar operators that play the role of fundamental
fields in the LGW theory that should provide an effective description
of the critical dynamics. In the effective model, no gauge fields are
considered.  The lattice Abelian-Higgs model with compact gauge fields
and unit-charge $N$-component scalar fields is an example of this type
of behavior~\cite{PV-19-AH3d,PV-19}.

At present, for 3D nonabelian gauge theories, no continuous transition
has been identified where the critical behavior can be conclusively
associated with stable charged FPs of the corresponding nonabelian
continuum field theory.  Models with SU($N_c$) and SO($N_c$) local
invariance have been numerically studied in
Refs.~\cite{BPV-19,BPV-20-su,BPV-20-on}, but in all cases gauge fields
were found to be not critical along the transition lines identified in
these models: the critical behavior could be explained in terms of
effective LGW models of the scalar order parameter, without gauge
fields.  Some hints of a new critical behavior have been reported for
SU($N_c$) gauge theories with scalar matter in the adjoint
representation~\cite{BFPV-21-3d}, but the role of gauge fields is not
yet clear.

In the following we consider the continuum SU($N_c$) gauge field theory
that corresponds to the lattice model, to check whether, and
under which conditions, charged FPs emerge. As in the lattice model, 
the fundamental fields 
are a complex matrix $\Phi^{af}(\bm{x})$ ($a=1,...,N_c$ and
$f=1,...,N_f$), and an SU($N_c$) gauge field $A_{\mu}^a(\bm{x})$.
The Lagrangian is
\begin{eqnarray}
{\cal L}&=& {1\over 4 g_0^2} {\rm Tr}\,F_{\mu\nu}^2 + {\rm Tr} [(D_\mu
    \Phi)^\dagger (D_\mu \Phi)] 
    \label{cogau} \\
      &+& {r\over 2}  \,{\rm Tr}\,\Phi^\dagger\Phi 
        + {u_0 \over 4} ({\rm Tr}\,\Phi^\dagger\Phi)^2 + {v_0\over 4} 
  {\rm Tr}\,(\Phi^\dagger\Phi)^2 \,,
  \nonumber
\end{eqnarray}
where $F_{\mu\nu} = \partial_\mu A_\nu -\partial_\nu A_\mu -i[A_\mu,
  A_\nu]$ and $D_{\mu, ab} = \partial_\mu\delta_{ab} -i t_{ab}^c
A_\mu^c$ where $t^c$ are the 
SU($N_c$) Hermitian generators in the
fundamental representation.  

To determine the nature of the transitions described by the continuum
SU($N_c$) gauge theory (\ref{cogau}), one studies the RG flow
determined by the $\beta$ functions of the model in the coupling
space.  Within the $\epsilon$-expansion framework, the RG flow close
to four dimensions is determined by the one-loop $\overline{\rm MS}$
$\beta$ functions.  Introducing the renormalized couplings $u$, $v$,
and $\alpha = g^2$, the corresponding $\overline{\rm MS}$ one-loop
$\beta$ functions read~\cite{CP-inprep}
\begin{eqnarray}
&&\beta_\alpha = -
  \epsilon \alpha + (N_f-22N_c)\,\alpha^2\,,
\label{betas}\\
&&\beta_u = -\epsilon u + (N_f N_c + 4) u^2   
+ 2 (N_f+N_c) u v + 3 v^2  
\nonumber\\
&&\;\; 
 - {18 \,(N_c^2 -1)\over N_c} \, u \alpha 
 + {27 (N_c^2 + 2)\over N_c^2}\,\alpha^2
 \,,
\nonumber\\
&&\beta_v = - \epsilon v + (N_f+N_c)v^2 + 6uv
- {18 \,(N_c^2 -1)\over N_c} \, v \alpha
\nonumber\\
&&\;\;
+ {27 (N_c^2 - 4)\over N_c}\,\alpha^2
\,,
  \nonumber
\end{eqnarray}
where $\epsilon\equiv 4-d$. The normalizations of the renormalized
couplings can be inferred from the above expressions.

Close to four dimensions, a stable FP occurs for $N_f > N_f^*$ with
$N_f^*= 375.4+O(\epsilon)$ for $N_c=2$, and $N_f^* =
638.9+O(\epsilon)$ for $N_c=3$. The stable FP for $N_f>N_f^*$ is
located in the region with positive values of $v$.  This can be also
inferred by considering the large-$N_f$ limit.  In this case the
$\beta$ functions (\ref{betas}) can be expressed in terms of
$\hat{u}\equiv N_f u$, $\hat{v}\equiv N_f v$, and $\hat{\alpha}\equiv
N_f \alpha$, as
\begin{eqnarray}
&&\beta_{\hat{\alpha}} = -\epsilon \hat{\alpha}  + \hat\alpha^2 \,,
  \label{largeNfbetas}\\
&&  \beta_{\hat{u}} = -\epsilon \hat{u} + N_c \hat{u}^2 +
       2 \hat{u}\hat{v} \,,  \nonumber\\
&&\beta_{\hat{v}} = - \epsilon \hat{v} + \hat{v}^2 \,,\nonumber
\end{eqnarray}
which have a stable FP for
\begin{eqnarray}
  \hat\alpha^* = \epsilon\,,\quad \hat{u}^* = 0, 
  \quad \hat{v}^* = \epsilon\,.
  \label{fpln}
\end{eqnarray}
Since the stable FP in the large-$N_f$ limit is located in the region
$v>0$, it should describe continuous transitions between the
disordered phase and the positive-$v$ Higgs phase discussed in
Sec.~\ref{posvcase}. Thus the, corresponding symmetry breaking pattern
should be that reported in Eq.~(\ref{glsymbrpat}).

We also note that the uncharged FP with vanishing gauge coupling
($\alpha=0$) is always unstable with respect to the gauge coupling,
since the stability matrix $\Omega_{ij} = \partial \beta_i/\partial
g_j$ has a a negative eigenvalue
\begin{equation}
\lambda_\alpha = 
\left. {\partial \beta_\alpha \over \partial \alpha} \right|_{\alpha=0}
= - \epsilon + O(\epsilon^2)\,.
\label{lambdares}
\end{equation}

\section{Predicted phase diagrams}
\label{phadia} 

In this section we sketch the phase diagrams using the theoretical
arguments presented in Sec.~\ref{gausym} and known results for
particular limiting cases.  We will always assume $N_f>1$, since for
$N_f=1$ the phase diagram of the model consists of a single
phase~\cite{OS-78,FS-79,DRS-80}.  The predictions will be checked
numerically for $N_c= 2$ and several values of $N_f$ in the next
section.  We mention that the phase diagram and critical behavior for
$v=0$ were investigated in Refs.~\cite{BPV-19,BPV-20-su}.

\subsection{Some particular cases}
\label{particase}

In the limit $\beta\to\infty$ the behavior of the system is determined
by the configurations minimizing the Hamiltonian.  As already
discussed in Sec.~\ref{gausym}, for $v < 0$ and $v > 0$, the relevant
configurations are different. For $N_f\ge N_c$ we expect two different
Higgs phases depending on the sign of $v$, while, for $N_f<N_c$, there
is one Higgs phase only for $v < 0$.  For positive values of $v$ the
system is disordered up to $\beta = \infty$.  We therefore expect a
first-order transition for $v=0$ and any $\gamma$.  This first-order
transition is the endpoint of a transition line (transition surface if
we also consider the parameter $\gamma$) for finite values of $\beta$.
Its behavior depends on $N_c$.  For $N_c=2$ the global symmetry for
$v=0$ is larger than for $v\not=0$~\cite{BPV-19,BPV-20-su}.
Therefore, for $N_c=2$ the finite-$\beta$ transition line between the
two different low-temperature phases is expected to run along the
$v=0$ axis.  This is not true for $N_c > 2$, where the transition line
between the different low-temperature phases converges to $v=0$ only
for $\beta\to\infty$.

In the limit $\gamma\to\infty$, the gauge variables $U_{{\bm x},\mu}$
are equal to the identity (strictly speaking, this is correct only in
the infinite-volume limit), apart from gauge transformations. Thus,
the scalar fields interact with Hamiltonian
\begin{eqnarray}
  H = - N_f \sum_{{\bm x},\mu}
{\rm Re}\,  {\rm Tr} \,\Phi_{\bm x}^\dagger
  \, \Phi_{{\bm x}+\hat{\mu}}^{\phantom\dagger}
+ {v\over 4} \sum_{\bm x}  {\rm Tr}\,(\Phi_{\bm x}^\dagger\Phi_{\bm x})^2\,,
\label{hfixedlengthgammainf}  
\end{eqnarray}
with global symmetry U($N_f$)$\otimes$U($N_c$).  For $v=0$ the
symmetry enlarges to O($M$) with $M=2N_fN_c$, so that continuous
transition should belong to the O($M$) vector universality class. The
behavior of model (\ref{hfixedlengthgammainf}) for $v\not=0$ can be
predicted by studying the RG flow of the LGW $\Phi^4$ theory with the
same global symmetry: continuous transitions are possible only if a
stable FP exists.  Results for $N_f=N_c$, the relevant case for the
chiral finite-temperature transition of the strong-interaction theory
in the massless quark limit, are reported in
Refs.~\cite{PW-84,BPV-03,PV-13}.  High-order 3D perturbative
schemes~\cite{PV-13} indicate the presence of a stable FP (with $v >
0$) only for $N_f=N_c=2$; no stable FPs are found for
$N_f=N_c>2$. Results for different $N_c$ and $N_f$ are presented in
Ref.~\cite{CP-04}. Stable FPs (again with $v>0$) exist for
sufficiently large $N_f>N_c$ ~\cite{CP-04}.  In particular, for
$N_c=2$ the analysis~\cite{CP-04} of five-loop $\epsilon$ expansions
shows that a 3D stable FP exists for $N_f\gtrsim 5$ [close to four
  dimensions, a stable FP exists only for $N_f>N_f^*$ with
  $N_f^*=18.4853 +O(\epsilon)$].

The FPs occurring for $\gamma=\infty$ are expected to be unstable with
respect to gauge interactions, as suggested by the RG analysis
reported in Sec.~\ref{sft}: as soon as $\gamma$ is finite (or $\alpha$
is positive in the notations of Sec.~\ref{sft}), the RG flow moves
away from the infinite-$\gamma$ FP.  However, for large values of
$\gamma$, the infinite-$\gamma$ FP may give rise to sizeable crossover
effects, somehow controlling a preasymptotic regime at phase
transitions.

\subsection{Phase diagrams for $N_c=2$}
\label{phdianc2}

The phase diagram of lattice SU(2) gauge theories differs from that of
models with $N_c>2$. This is related to the presence for $v = 0$ of a
larger global symmetry: the theory is invariant under the
Sp($N_f$)$/\mathbb{Z}_2$ group, which is larger than the U($N_f$)
symmetry group of the model for generic $v\neq 0$.  This implies that
transitions between the different low-temperature phases discussed in
Sec.~\ref{gausym} must be located within the plane $v=0$ of the
$\beta$-$v$-$\gamma$ phase diagram.

\subsubsection{The case $N_f=N_c=2$.}
\label{nf2nc2}

\begin{figure}[tbp]
\includegraphics[width=0.95\columnwidth, clip]{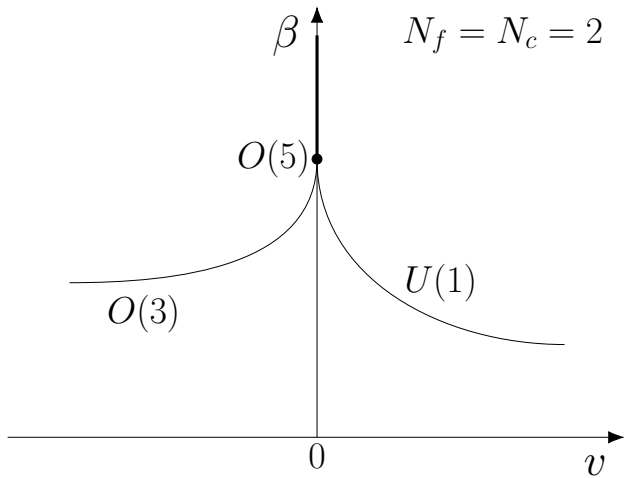}
\caption{A sketch of the phase diagram expected for $N_f= N_c=2$ in
  $\beta$-$v$ planes at finite $\gamma\ge 0$. There are two transition
  lines for $v\not=0$, meeting at a multicritical point $(v =0, \beta
  = \beta_{mc})$ with O(5) simmetry. Continuous transitions would
  belong to the O(3) and $\hbox{U(1)}=\hbox{SO(2)}$ vector
  universality classes for $v < 0$ and $v > 0$, respectively.
  First-order transitions are expected on the line $(v =0, \beta >
  \beta_{mc})$.  }
  \label{phdianf2nc2}
\end{figure}

A sketch of the expected phase diagram for $N_f=N_c=2$ for a fixed
value of $\gamma$ is reported in Fig.~\ref{phdianf2nc2}.  We expect it
to qualitatively apply to any finite $\gamma\ge 0$, except in the
$\gamma\to\infty$ limit, as discussed in Sec.~\ref{particase}.

For $v<0$, as discussed in Sec.~\ref{negvcase}, we expect the model to
behave as the CP$^1$ model, so that continuous transitions should
belong to the O(3) vector universality class.  For $v > 0$, instead,
as discussed in Sec.~\ref{posvcase}, we expect a transition line where
the U(1) degrees of freedom condense.  The two lines are expected to
meet at a multicritical point at $v=0$, where the global symmetry
enlarges to Sp(2)/${\mathbb Z}_2$=SO(5), due to the pseudoreality of
the SU(2) group; see, e.g.,
Refs.~\cite{Georgi-book,AY-94,DP-14,WNMXS-17} for a discussion in the
continuum theory and Refs.~\cite{BPV-19,BPV-20-su} for the lattice
case.  Therefore, the critical behavior should belong to the O(5)
vector universality class.  At the multicritical point, the two order
parameters $Q$ and $Y$ defined in Sec.~\ref{obsfss} both show
long-range order. Indeed, at the multicritical point one can define a
five-component real order parameter~\cite{BPV-19,BPV-20-su} that
combines $Q$ and $Y$:
\begin{eqnarray}
  &&\varphi_{\bm x}^{(k)} = \sum_{fg} \sigma^k_{fg} Q_{\bm x}^{fg}\,,\quad
  k=1,2,3\,,\label{O3op}\\
  &&\varphi_{\bm x}^{(4)} + i \varphi_{\bm x}^{(5)}  = 
  {1\over 2} \sum_{fg} \epsilon_{fg} Y_{\bm x}^{fg}={\rm det}\,\Phi\,,\label{U1op}
  \end{eqnarray}
where $\sigma^k$ are the Pauli matrices.

Note that multicritical points arising from the competition of O(3)
and U(1) order parameters do not generally lead to a multicritical
behavior with an enlarged SO(5) symmetry, as discussed in
Refs.~\cite{CPV-03,HPV-05}. In the case at hand, this occurs because
the model for $v=0$ is exactly invariant under the larger group
Sp(2)/${\mathbb Z}_2=$SO(5).

Close to the multicritical point, the free energy
can be written as~\cite{CPV-03,HPV-05,PV-02}
\begin{equation}
F_{\rm sing}= t^{3\nu} f_{\rm mc}(v t^{-\phi_T}) ,
\label{freen}
\end{equation}
where $t \sim \beta - \beta_c(v=0)$.  In
particular,~\cite{BPV-19,BPV-20-su} $\beta_c(v=0) = 2.68885(5)$ and
$\beta_c(v=0) = 1.767(1)$ for $\gamma=0$ and $\beta\gamma=2$
(i.e.,$\gamma\approx 1.13$), respectively.  Here $\nu$ is the O(5)
correlation-length exponent, $\nu=0.779(3)$ (Ref.~\cite{HPV-05}), and
$\phi_T>0$ is the crossover exponent associated with the RG dimension
$y_{2,2}$ of the relevant spin-2 quadratic perturbation at the O(5)
vector FP.  This is given by $\phi_T = y_{2,2} \,\nu$
with~\cite{CPV-03} $y_{2,2} = 1.832(8)$, thus $\phi_T = 1.427(8)$.
Since the transition lines $\beta_c(v)$ for $v>0$ and $v<0$ correspond
to constant values of the argument of the scaling function $f_{\rm
  mc}$, from the scaling behavior (\ref{freen}) it follows that
\begin{equation}
   |\beta_c(v) - \beta_c(v=0)|\sim |v|^{\zeta}\,,\qquad \zeta=\phi_T^{-1}<1\,.
  \label{lineappro}
\end{equation}
This implies that the $v>0$ and $v<0$ transition lines must approach
the $v=0$ axis tangentially.

It is interesting to compare the 3D phase diagram with the one
expected for finite-temperature 3D quantum systems, i.e., for the
analogous lattice SU(2) gauge model defined on a ($3+1$)-dimensional
lattice in which the number $L_t$ of sites in the fourth direction is
fixed.  In this case, in the absence of matter fields, we have also a
finite-$\gamma$ ${\mathbb Z}_2$ transition associated with the
breaking of the center symmetry of the SU(2) gauge group.  Such a line
may also be present in the theory with scalar fields for small values
of $\beta$, since, at small $\beta$, the integration of the scalar
fields can only give rise to a renormalization of the gauge coupling.

\subsubsection{The case $N_f>N_c=2$.}
\label{nfgtnc2}

\begin{figure}[tbp]
\includegraphics[width=0.95\columnwidth, clip]{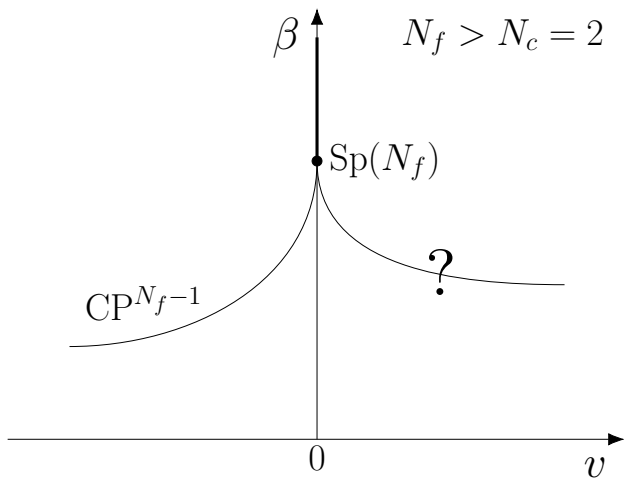}
\caption{A sketch of the phase diagram expected for $N_f>N_c=2$ for
  fixed values of $\gamma\ge 0$. For $v < 0$, $\gamma$ should not play
  any role and the transition line should of first order or belong to
  the CP$^{N_f-1}$ universality class, if it exists. For $v > 0$, the
  nature of the transition might depend on $\gamma$ for sufficiently
  large values of $N_f$. For $v=0$ we have a first-order line ending
  at a first-order multicritical point.  }
  \label{phdianfgtnc2}
\end{figure}

Let us now consider the case $N_f>N_c=2$. The expected phase diagram
is shown in Fig.~\ref{phdianfgtnc2}. Also in this case we have two
different Higgs phases for $\beta\to \infty$, characterized by
different global symmetry breaking patterns, and an enlargement of the
symmetry for $v=0$.

For $v<0$ the transition should behave as in the CP$^{N_f-1}$ model.
Generically, we expect a first-transition line except, possibly, for
small values of $N_f$ (we recall that the question of the existence of
continuous transitions in CP$^{N_f-1}$ models is stll debated
\cite{NCSOS-11,NCSOS-13,PV-19}).  For $v < 0$, we do not expect
$\gamma$ to be relevant. Indeed, the field-theory analysis of
Sec.~\ref{sft} shows that the RG flow for $v < 0$ does not have stable
FPs.  Thus, no charged critical behavior is expected.

For $v > 0$, we expect a transition line associated with the symmetry
breaking pattern (\ref{glsymbrpat}). The nature of the transition is,
however, not clear, since for $N_f > N_f^\star$, the field-theory RG
flow has a stable FP, see Sec.~\ref{sft}, indicating that gauge modes
can become critical and change the critical behavior. Therefore, {\em
  a priori} two different types of critical behavior can occur.  For
$N_f < N_f^\star$, $\gamma$ should not play any role and the gauge
model should behave as the effective matrix model obtained by
integrating the gauge degrees of freedom (see App.~\ref{AppA}). The
numerical results of the next section indicate that the transition
line is of first order.  For $N_f \ge N_f^\star$, instead, one might
have two different regimes, depending on $\gamma$. For small $\gamma$,
the effective matrix model describes the critical behavior, while for
large values of $\gamma$ a new critical behavior sets in, controlled
by the field-theory charged FP.

As it occurs for $N_f=2$, for $v=0$ the symmetry enlarges to
Sp($N_f$)/${\mathbb Z}_2$.  Thus, we have a multicritical point for
$v=0$ (LGW arguments predict the transition to be of first-order for
any $N_f\ge 3$ \cite{BPV-19,BPV-20-su}) and a first-order transition
line, extending from the multicritical point to $\beta = \infty$ along
the $v=0$ axis.

\subsection{Phase diagrams for $N_c\ge 3$}
\label{ncge3}

We now sketch the possible phase diagrams for $N_c\ge 3$. We must
distinguish three cases, i.e., $N_f<N_c$, $N_f=N_c$ and $N_f>N_c$.

\begin{figure}[tbp]
\includegraphics[width=0.95\columnwidth, clip]{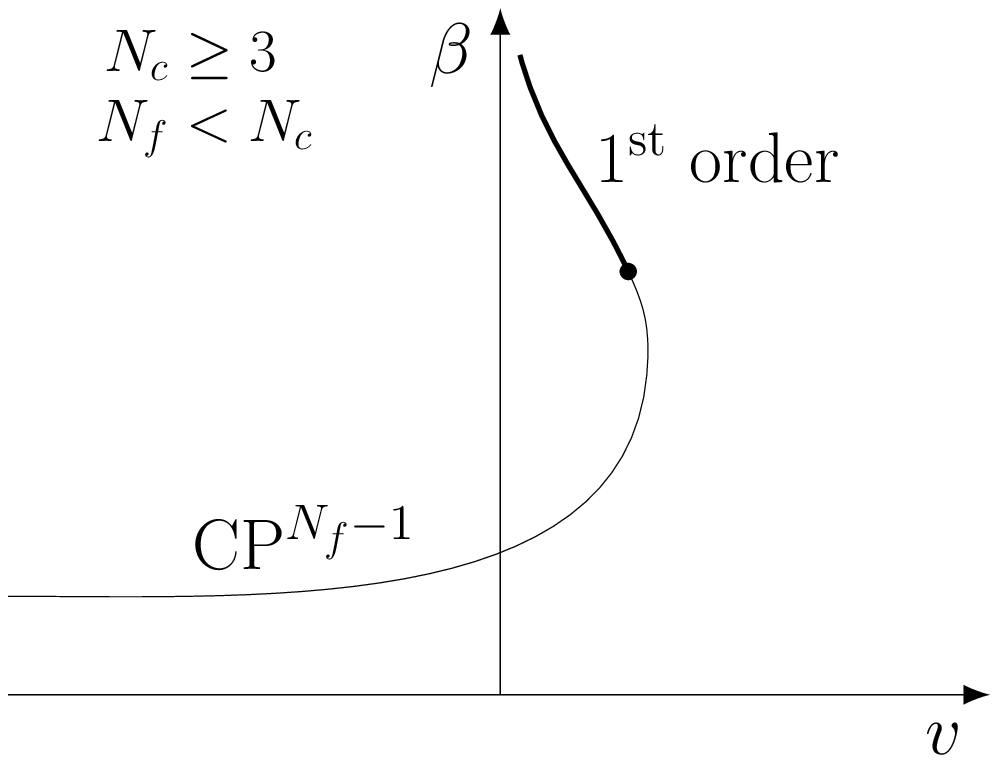}
\caption{A sketch of the phase diagram at fixed $\gamma\ge 0$ expected
  for $N_f<N_c$, $N_c\ge 3$. For values of $N_f$ for which there is no
  CP$^{N_f-1}$ universality class, the whole line corresponds to first
  order transitions. }
  \label{phdianfltncge3}
\end{figure}

In Fig.~\ref{phdianfltncge3} we show the expected phase diagram for
$N_f<N_c$ and $N_c\ge 3$.  For $v<0$, the behavior is independent of
$N_c$ and thus the high-temperature disordered phase and
low-temperature Higgs phase are separated by a transition line where
the system behaves as the CP$^{N_f-1}$ model. In particular, for
$N_f=2$, transitions may be continuous, in the O(3) vector
universality class. The results of Refs.~\cite{BPV-19,BPV-20-su}
indicate that this line intersects the $v=0$ axis at a finite $\beta$
value.  Presumably, it enters the $v>0$ half-plane. However, since for
$v$ large enough the system is disordered for any $\beta$, the curve
should bend and approach $v=0$ as $\beta\to \infty$. Note, that for
large $\beta$, transitions should be of first order, hence a
tricritical point should be present, if the transitions are continuous
for $v < 0$ (this is the expected behavior for $N_f=2$).

\begin{figure}[tbp]
\includegraphics[width=0.95\columnwidth, clip]{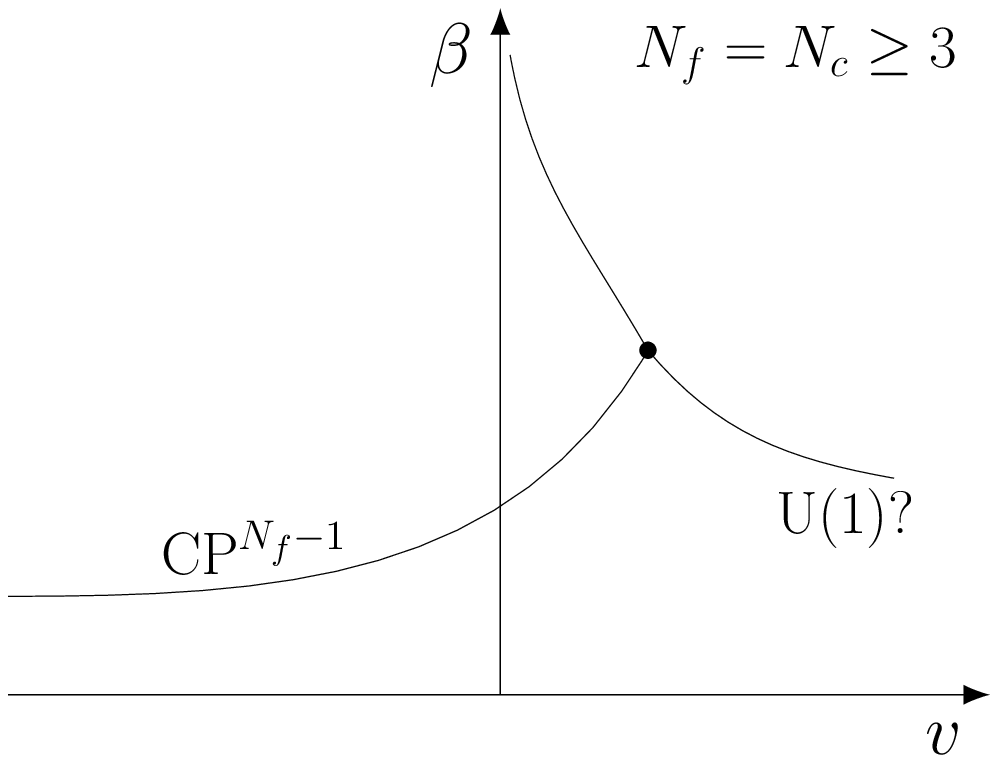}
\caption{A sketch of the phase diagram at fixed $\gamma\ge 0$ expected
  for $N_f=N_c\ge 3$.}
  \label{phdianfeqncge3}
\end{figure}

\begin{figure}[tbp]
\includegraphics[width=0.95\columnwidth, clip]{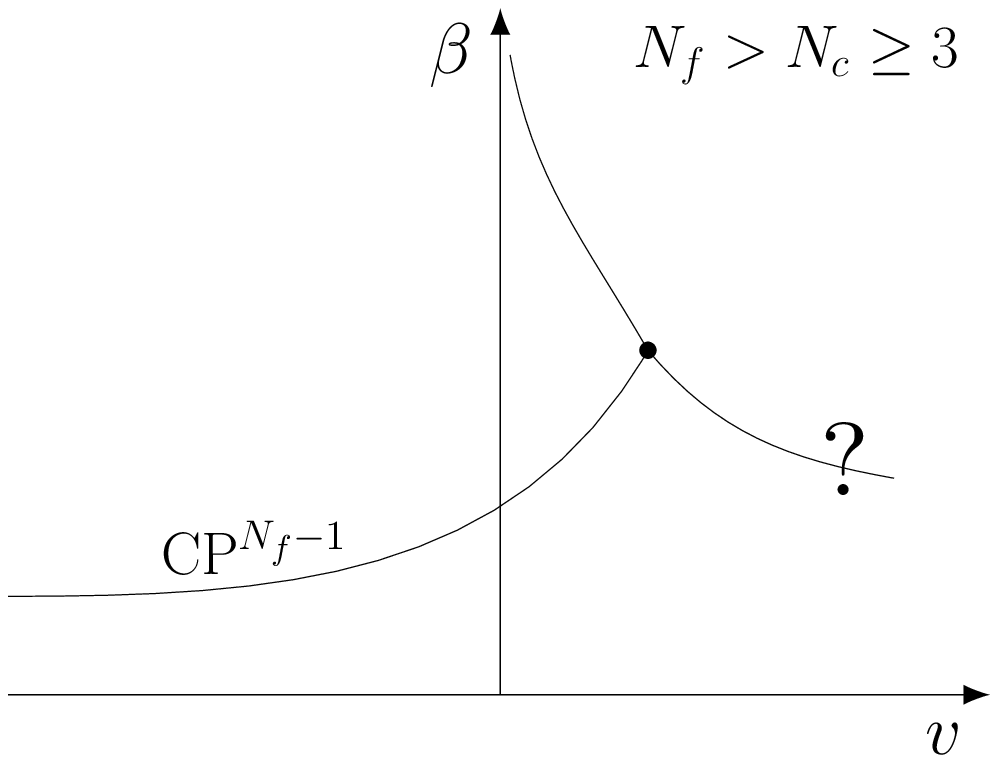}
\caption{A sketch of the phase diagram at fixed $\gamma\ge 0$ expected
  for $N_f>N_c\ge 3$.}
  \label{phdianfgtncge3}
	\end{figure}

A possible phase diagram for $N_c=N_f$ is shown in
Fig.~\ref{phdianfeqncge3}, while the case $N_c < N_f$ is reported in
Fig.~\ref{phdianfgtncge3}. The qualitative behavior in these two cases
should be similar to that observed for $N_c =2$. The only difference
is the absence of an enlarged symmetry for $v=0$, so that the $v=0$
axis does not play any particular role. Therefore, the multicritical
point, on whose nature we have no prediction, will be a generic point
with $v \not=0$. Analogously, the first-order transition line that
separates the two low-temperature Higgs phases will be a generic line
in the $\beta-v$ plane for each value of $\gamma$.  The considerations
we made on the nature of the transition lines, but not of the
multicritical points, in Sections~\ref{nf2nc2} and \ref{nfgtnc2} do
not depend on $N_c$ and also apply here.

\section{Numerical analyses for $N_c=2$}
\label{numresnc2}

In this section we present some numerical results for $N_c=2$, to
check the phase diagrams put forward in Sec.~\ref{phdianc2}. Some
technical details on the MC simulations are reported in
App.~\ref{mcsim}.

\subsection{Results for $N_f=N_c=2$}
\label{resnc2nf2}

To verify the phase diagram sketched in Fig.~\ref{phdianf2nc2}, and in
particular the existence of the O(3) and U(1) transition lines meeting
at the O(5) multicritical point located at $v=0$ axis and
~\cite{BPV-19,BPV-20-su} $\beta_c = 2.68885(5)$, we performed
numerical simulations for $v= 1$ and $v=-1$. As the parameter $\gamma$
should not play any role, we only performed simulations for $\gamma =
0$.

\begin{figure}[tbp]
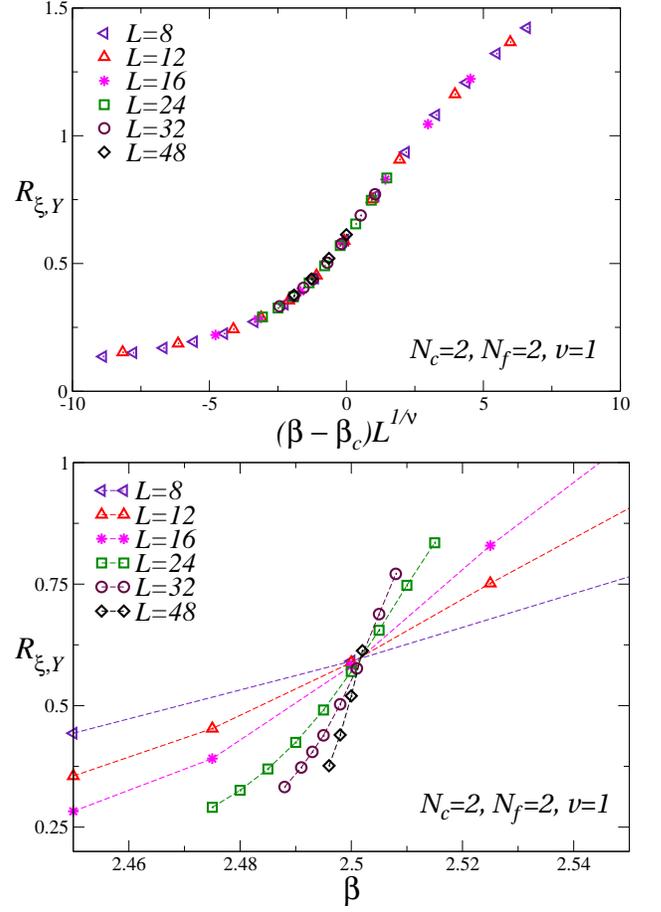

    \includegraphics[width=0.95\columnwidth, clip]{sun2c2fv1_detrxiscaling_beta.eps}
    \includegraphics[width=0.95\columnwidth, clip]{sun2c2fv1_detrxi_beta.eps}
  \caption{Data of $R_{\xi,Y}$ for $N_c=N_f=2$, $\gamma=0$ and $v=1$,
    versus $\beta$ (bottom) and versus $(\beta-\beta_c)L^{1/\nu}$
    (top). We use the XY critical exponent $\nu=0.6717$ and our best
    estimate $\beta_c=2.502$ of the critical point. The excellent
    collapse of the data (top panel) demonstrates that the transition
    belongs to the XY universality class.}
\label{detnc2nf2v1}
\end{figure}

\begin{figure}[tbp]
  \includegraphics[width=0.95\columnwidth, clip]{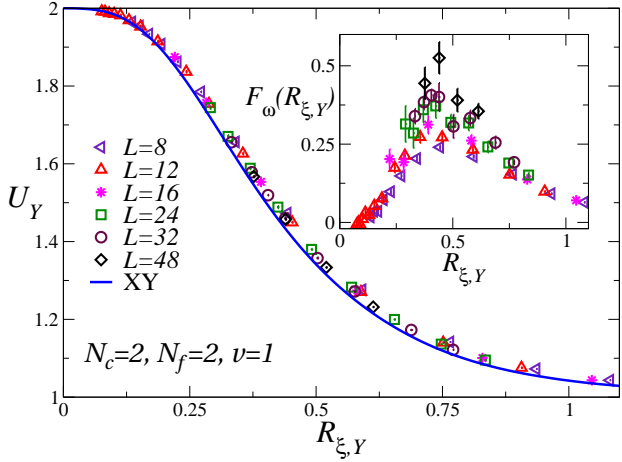}
  \caption{Estimates of $U_Y$ versus $R_{\xi,Y}$ for $N_c=N_f=2$,
    $\gamma=0$ and $v=1$. The continuous line is the XY universal
    curve $F(R_\xi)$ (taken from Ref.~\cite{BPV-21-coAH}). Estimates
    of $L^\omega[U_Y-F(R_{\xi,Y})]$ versus $R_{\xi,Y}$, using the XY
    correction-to-scaling exponent $\omega=0.789$, are reported in the
    inset. The data show a reasonable scaling behavior, which is
    definitely consistent with the scaling behavior (\ref{udiff}).  }
\label{detnc2nf2urxiv1}
\end{figure}

\begin{figure}[tbp]
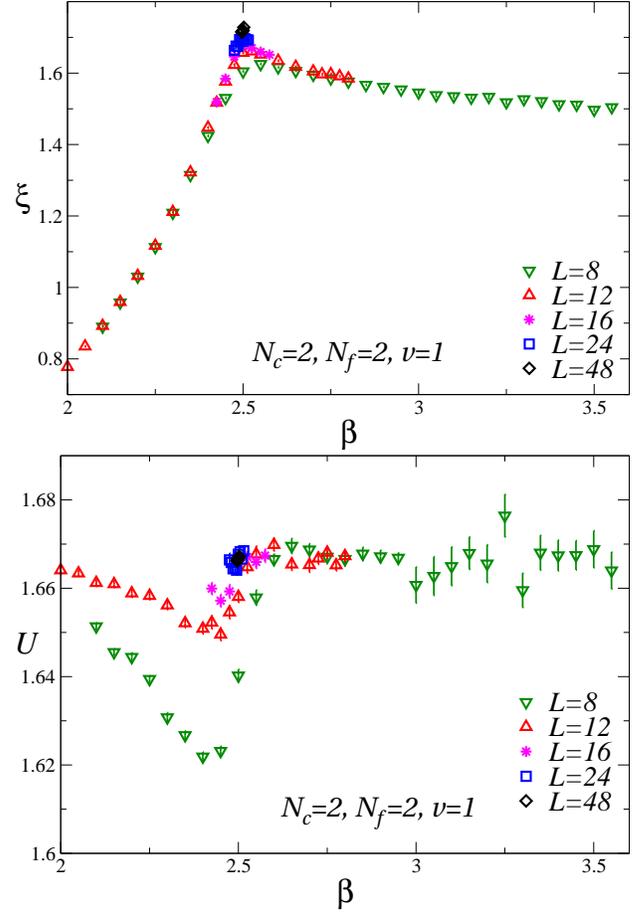

  \includegraphics[width=0.95\columnwidth, clip]{sun2c2fv1_xi_beta.eps}
  \includegraphics[width=0.95\columnwidth, clip]{sun2c2fv1_U_beta.eps}
  \caption{Data of $\xi$ (top) and $U$ (bottom) for $N_c=N_f=2$,
    $\gamma=0$ and $v=1$, as obtained from the correlations of the
    bilinear operator $Q_{\bm x}$. They clearly show that the
    correlations of the bilinear operator $Q$ do not become critical
    for $v>0$, as expected. }
\label{Qnc2nf2v1}
\end{figure}

For $v=1$, the estimates of the order parameter $Y_{\bm x}^{fg}$
defined in Eq.~(\ref{Ydet}), are reported in Figs.~\ref{detnc2nf2v1}
and \ref{detnc2nf2urxiv1}. The data confirm the existence of a
continuous transition at $\beta \approx 2.50$, which belongs to the
U(1), or XY, universality class. Indeed, if we fit the data using the
XY estimate $\nu= 0.6717(1)$ (see
Refs.~\cite{CHPV-06,Hasenbusch-19,CLLPSSV-20,PV-02}) we obtain
$\beta_c = 2.502(1)$ and an excellent collapse of the data (upper
panel of Fig.~\ref{detnc2nf2v1}). The best evidence that the
transition belongs to the XY universality class is provided by the
plot of $U_Y$ versus $R_{\xi,Y}$. Data approach the asymptotic
universal curve $F(R_\xi)$ corresponding to the XY universality class
(the curve is taken from the appendix of
Ref.~\onlinecite{BPV-21-coAH}).  Moreover, the approach to the
universal XY curve, see the inset of Fig.~\ref{detnc2nf2urxiv1}, is
consistent with the expected FSS scaling behavior
\begin{equation}
  U(L,R_{\xi,Y}) - F(R_{\xi,Y}) \approx L^{-\omega}
  F_{\omega}(R_{\xi,Y})\,,
  \label{udiff}
\end{equation}
where $\omega$ is the leading scaling-correction exponent and
$F_{\omega}(R_\xi)$ is a scaling function that is universal apart from
a multiplicative factor.  If we use the XY
estimate~\cite{Hasenbusch-19} $\omega=0.789(4)$, we observe a
reasonable scaling, again confirming that the transition is related to
the breaking of the U(1) symmetry. The SU($N_f$) symmetry is unbroken
and the indeed, correlations of the bilinear operator $Q$ are not
critical (but still nonanalytic) for $v>0$, as expected, see
Fig.~\ref{Qnc2nf2v1}.

\begin{figure}[tbp]
  \includegraphics[width=0.95\columnwidth, clip]{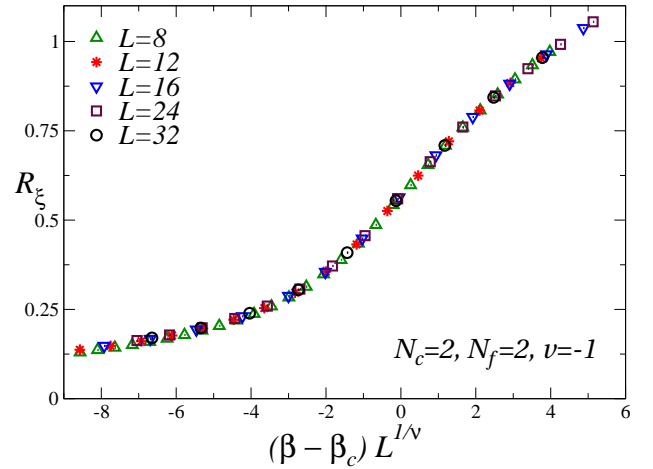}
  \caption{Plot of $R_{\xi}$ for $N_c=N_f=2$, $\gamma=0$ and $v=-1$,
    versus $(\beta-\beta_c)L^{1/\nu}$, using the O(3) critical
    exponent \cite{Hasenbusch-19} $\nu=0.71164$ and our best estimate
    $\beta_c=2.561$ of the critical point.  The excellent collapse of
    the data supports the O(3) critical behavior.}
\label{sunc2nf2rxivm1}
\end{figure}

\begin{figure}[tbp]
  \includegraphics[width=0.95\columnwidth, clip]{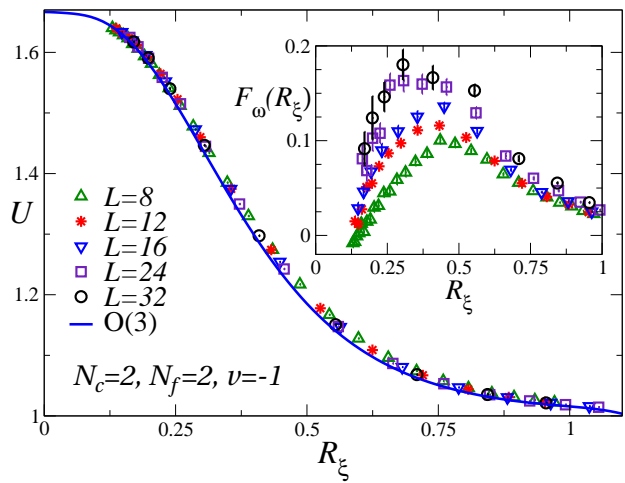}
  \caption{$U$ versus $R_{\xi}$, as obtained from the correlations of
    the bilinear operator $Q$, for $N_c=N_f=2$, $\gamma=0$ and
    $v=-1$. The data appear to approach the universal curve
    corresponding to the O(3) universality class ~\cite{BPV-21-coAH}.
    In the inset we report $F_\omega(R_\xi)\equiv L^\omega[U
      -F(R_{\xi})]$ versus $R_{\xi}$, using the O(3)
    correction-to-scaling exponent~\cite{Hasenbusch-20}
    $\omega=0.759$.  The data, in particular those for the largest
    available lattice sizes, show a reasonably good scaling.}
\label{sunc2nf2urxivm1}
\end{figure}

In Figs.~\ref{sunc2nf2rxivm1} and \ref{sunc2nf2urxivm1} we report
results for $v=-1$. In this case, the order parameter $Q_{\bm x}^{fg}$
defined Eq.~(\ref{qdef}) is critical, signalling the breaking of the
SU(2) symmetry and therefore the presence of a transition that belongs
to the CP$^1$ or O(3) universality class.  In
Fig.~\ref{sunc2nf2rxivm1} we report a scaling plot of $R_\xi$ using
~\cite{Hasenbusch-20} the O(3) estimate $\nu= 0.71164(10)$ (accurate
estimates of the O(3) exponents can be found in
Refs.~\cite{Hasenbusch-20,
  Chester-etal-20-o3,KP-17,HV-11,PV-02,CHPRV-02, GZ-98}) and the
estimate of the critical temperature $\beta_c=2.561(1)$, obtained by
performing biased fits of the data, in which $\nu$ was fixed to the
O(3) value. The agreement is excellent. As before, we also considered
$U$ versus $R_\xi$. Data fall on top of the O(3) curve (it is reported
in the appendix of Ref.~\cite{BPV-21-coAH}) with small corrections
that are consistent with Eq.~(\ref{udiff}) and the O(3) value of the
scaling-correction exponent,~\cite{Hasenbusch-20} $\omega=0.759(2)$.
We also mention that correlations of the operator $Y_{\bm x}$ are not
critical, as expected.

In conclusion, our numerical results confirm the discussion of
Sec.~\ref{phdianc2}, and are fully consistent with the phase diagram
reported in Fig.~\ref{phdianf2nc2}. We expect the same qualitative
behavior for any finite inverse gauge coupling $\gamma>0$.

\subsection{Results for $N_f>N_c=2$}
\label{resnc2nfgt2}

We now present some numerical results for two large values of $N_f$,
$N_f=20$ and $N_f=40$, to check whether the lattice model develops a
critical behavior that can be associated with the charged FP of the
corresponding SU($N_c$) gauge field theory.  As we have discussed in
Sec.~\ref{sft}, the charged FP is expected to be present only if the
gauge fields develop a critical dynamics.  Therefore, we expect such a
behavior for nonvanishing values of $\gamma$.  For $\gamma = 0$, the
gauge fields can be integrated out and one obtains an effective scalar
model for the two order parameters, whose critical behavior should be
well described within the standard LGW approach without gauge fields,
see App.~\ref{AppA}.

\begin{figure}[tbp]
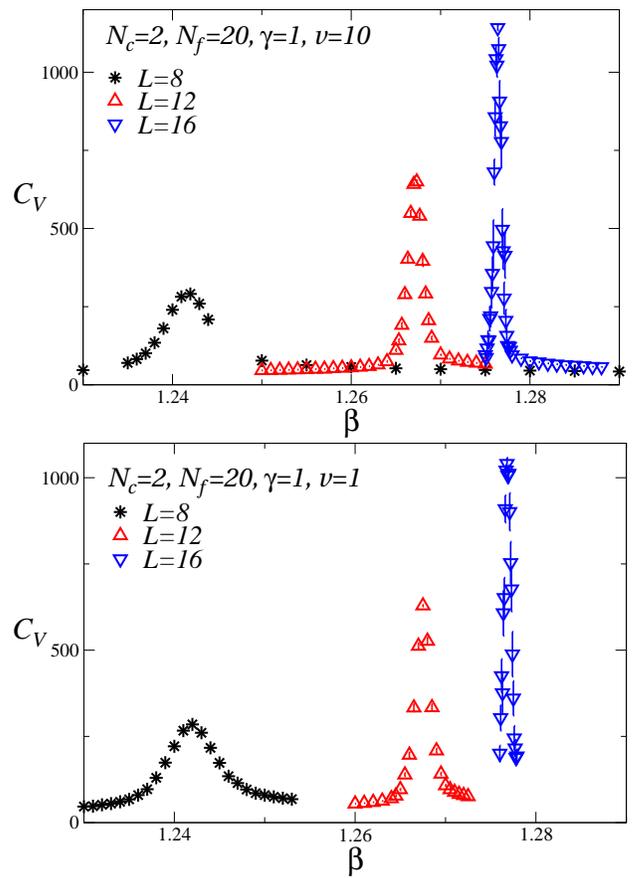

\includegraphics[width=0.95\columnwidth, clip]{sun2c20fg1v10_C_beta.eps}
\includegraphics[width=0.95\columnwidth, clip]{sun2c20fg1v1_C_beta.eps}
\caption{Data for the specific heat $C_V$ for $N_c=2$, $N_f=20$,
  $\gamma=1$, $v=10$ (top) and $v=1$ (bottom). The apparent divergence
  of $C_V$ with increasing $L$ supports a first-order transition.  }
\label{cvnc2nf20v1v10}
\end{figure}

\begin{figure}[tbp]
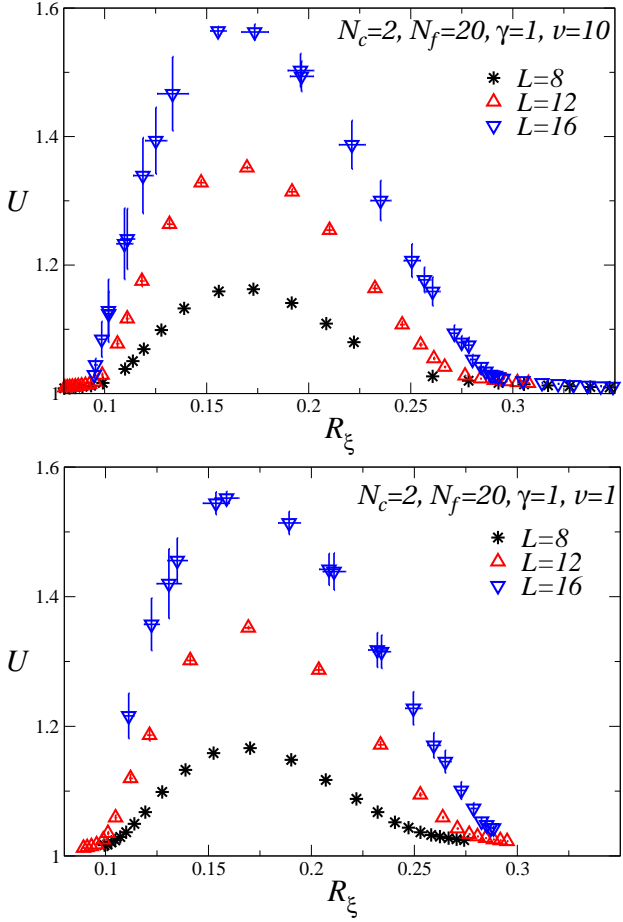

\includegraphics[width=0.95\columnwidth, clip]{sun2c20fg1v10_U_rxi.eps}
\includegraphics[width=0.95\columnwidth, clip]{sun2c20fg1v1_U_rxi.eps}
\caption{The Binder parameter $U$ versus the ratio $R_\xi$ for
  $N_c=2$, $N_f=20$, $\gamma=1$, $v=10$ (top) and $v=1$ (bottom). Data
  do not of converge and the maximum of the Binder parameter $U$
  increases with increasing $L$, as expected for a first-order
  transition (see the discussion at the end of Sec.~\ref{obsfss}).  }
\label{nc2nf20v1v10}
\end{figure}

\begin{figure}[tbp]
    \includegraphics[width=0.95\columnwidth, clip]{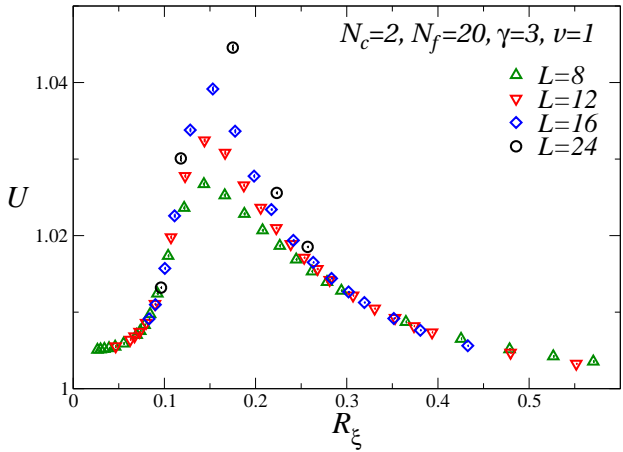}
    \caption{The Binder parameter $U$ versus the ratio $R_\xi$ for
      $N_c=2$, $N_f=20$, $v=1$, and $\gamma=3$. No scaling is
      observed, indicating that the transition is of first order. }
\label{nc2nf20v1g3}
\end{figure}

For $N_f = 20$, we have performed simulations for two values of
$\gamma$, choosing $\gamma = 1$ and 3.  For $\gamma = 1$, we have also
studied the $v$ dependence, considering $v = 1$ and $v=10$.  Results
for $\gamma = 1$ depend only weakly on $v$ and indeed, we find that
both models undergo a transition for a similar values of $\beta$,
$\beta_c\approx 1.28$. The results for the specific heat $C_V$ and 
the Binder parameter $U$ are shown in Figs.~\ref{cvnc2nf20v1v10}
and~\ref{nc2nf20v1v10}, respectively.  They are consistent with a first-order
transition: we do not observe scaling when $U$ is plotted against
$R_\xi$ and the maximum of $U$ increases with $L$.  Also the data for
$\gamma=3$ and $v=1$ favor a first-order transition, see
Fig.~\ref{nc2nf20v1g3}, at $\beta_c\approx 1.16$. The transition is
weaker than that observed for $\gamma=1$, and indeed larger lattices
are needed to observe the emergence of the typical features of
first-order transitions.  This is not unexpected, since the transition
may become continuous for $\gamma\to\infty$, controlled by the stable
FP of the matrix model (\ref{hfixedlengthgammainf}), see
Sec.~\ref{particase}.

As no evidence for a charged FP was found for $N_f=20$, we decided to
study the model for an even larger number of flavors.  We chose
$N_f=40$ and performed simulations for $\gamma=0$ and 1, and also for
the matrix model obtained in the limit $\gamma=\infty$ (it amounts to
setting $U_{{\bm x},\mu} = 1$ on every link). As $v$ does not play a
role, we always fixed $v = 1$.

\begin{figure}[tbp]
    \includegraphics[width=0.95\columnwidth, clip]{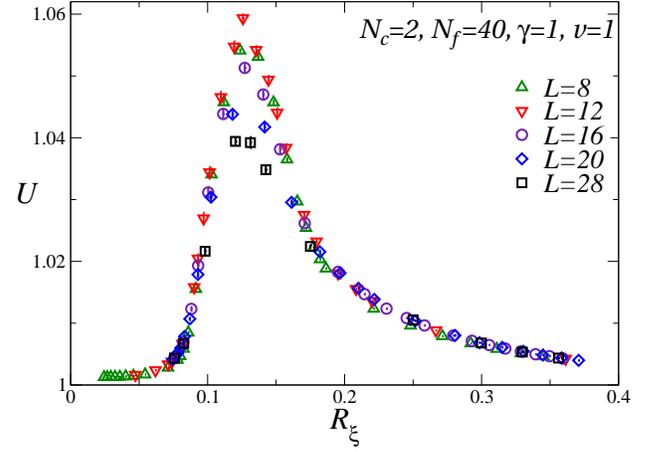}
    \caption{The Binder parameter $U$ versus the ratio $R_\xi$ for
      $N_c=2$, $N_f=40$, $v=1$, and $\gamma=1$. The data appear to
      converge to a scaling curve with increasing $L$, consistently
      with a continuous transition.  }
\label{nc2nf40v1g1}
\end{figure}

For $\gamma = 0$, we observe a very strong first-order transition at
$\beta_c \approx 1.2$. Already on small lattices, there are long-living
metastable states and we are not able to thermalize the system for
$L\gtrsim 12$. There is apparently no FP in the model in which gauge
fields are integrated out. To detect the possible presence of a
charged FP, we performed simulations for a finite value of $\gamma$,
choosing $\gamma = 1$. Results corresponding to $8\le L \le 28$ are
fully consistent with a continuous transition at $\beta_c\approx
1.18$.  First, the specific heat is apparently bounded---its maximum
does not increase with $L$. Second, the plot of the Binder parameter
versus $R_\xi$, see Fig.~\ref{nc2nf40v1g1}, shows a reasonably good
scaling.  In particular, the maximum of the Binder parameter does not
increase with $L$. On the contrary, it apparently decreases with
increasing sizes (we find $U_{\rm max} \approx 1.06$,1.04 for $L=12$
and 28, respectively), a phenomenon that is not consistent with a
first-order transition. The strong peak in the Binder parameter can be
interpreted as a crossover effect, due to the first-order transition
line that is expected to be present for smaller values of $\gamma$ and
that ends in the transition point at $\beta_c \approx 1.2$, $\gamma =
0$.

\begin{figure}[tbp]
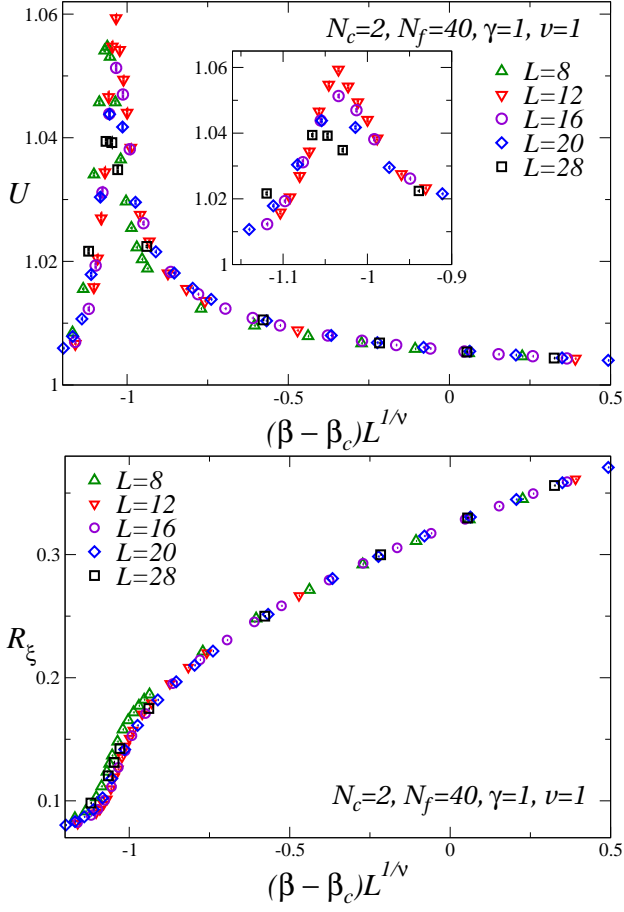

 \includegraphics[width=0.95\columnwidth, clip]{sun2c40fg1v1_U_scaling.eps}
 \includegraphics[width=0.95\columnwidth, clip]{sun2c40fg1v1_rxi_scaling.eps}
\caption{Binder paramer $U$ (top) and correlation-length ratio $R_\xi$
  (bottom) versus $X = (\beta - \beta_c) L^{1/\nu}$ for $N_c=2$,
  $N_f=40$, $v=1$, and $\gamma=1$.  We set $\beta_c = 1.1864$ and $\nu
  = 0.74$.  The inset in the upper panel gives a more detailed view of
  the behavior of the Binder parameter for $-1.2\le X\le -0.8$, the
  values of $X$ where $U$ has a peak.  }
\label{URxi-X-Nf40}
\end{figure}

\begin{figure}[tbp]
\includegraphics[width=0.95\columnwidth, clip]{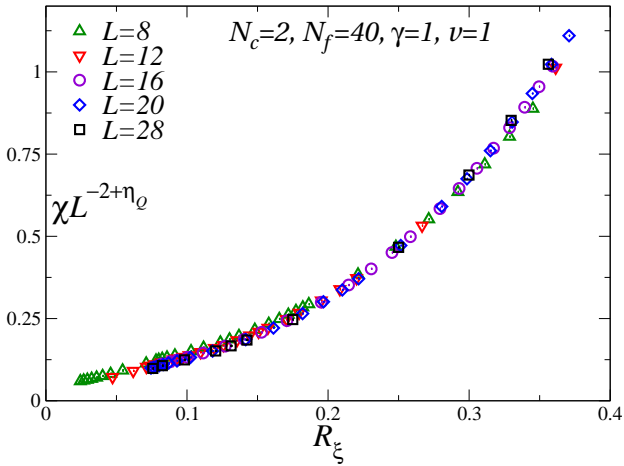}
\caption{Estimates of $L^{\eta_Q-2} \chi$ versus $R_\xi$, 
for $N_c=2$, $N_f=40$, $v=1$, and $\gamma=1$.
We set $\eta_Q = 0.89$.
}
\label{chi-X-Nf40}
\end{figure}

\begin{figure}[tbp]
    \includegraphics[width=0.95\columnwidth, clip]{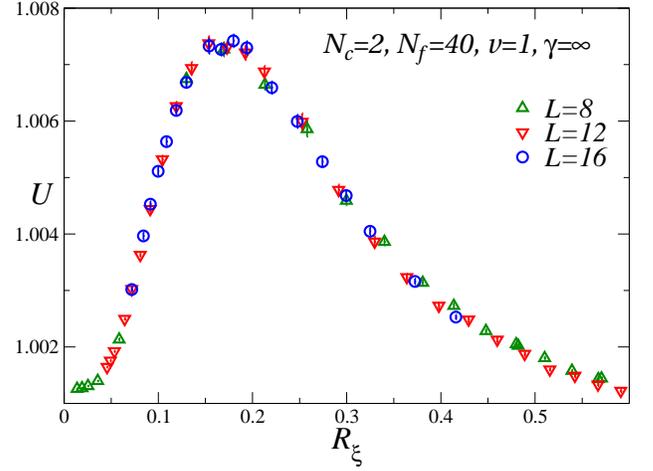}
    \caption{The Binder parameter $U$ versus the ratio $R_\xi$ for
      $N_c=2$, $N_f=40$, $v=1$, and $\gamma=\infty$ (ungauged matrix
      model).  }
\label{URxi-Nf40-gammainf}
\end{figure}

As the transition for $\gamma = 1$ is apparently continuous, it is
interesting to determine the corresponding critical exponents.  The
exponent $\nu$ has been determined by fitting $R_\xi$ to $f(X)$, with
$X = (\beta-\beta_c)L^{1/\nu}$. We have parameterized the function
$f(X)$ with an order-$n$ polynomial (stable results are obtained for
$n\gtrsim 15$). We have performed several fits, including each time
only data satisfying $L\ge L_{\rm min}$ (we used $L_{\rm min} =
8,12,16$).  Moreover, as corrections appear to be stronger in the
region where $U$ has a peak, we also investigated how results change
if only data satisfying $R_{\xi} \ge 0.20$ are considered.  The
results of these analyses are consistent with
\begin{equation}
\beta_c = 1.1864(1)\,,\qquad \nu = 0.74(2)\,.
\label{estimates-nuNf40}
\end{equation}
In Fig.~\ref{URxi-X-Nf40} we report the plots of $U$ and $R_\xi$
versus $X$, using the estimates (\ref{estimates-nuNf40}). We observe
good scaling, except for $X \lesssim -0.8$, where $U$ has a peak.
Note that $\nu > 2/3$, and thus the result is consistent with a finite
specific heat at the transition for $L \to \infty$.  We have also
estimated the exponent $\eta_Q$ that characterizes the behavior of the
susceptibility, $\chi \sim L^{2-\eta_Q}$ at the critical point. To
estimate $\eta_Q$ we have fitted $\log \chi$ to $(2-\eta_Q) \log L +
g_\chi(R_\xi)$, using a polynomial parametrization for the function
$g(x)$. We find
\begin{equation}
\eta_Q = 0.89(3).
\end{equation}
Scaling is excellent, as shown in Fig.~\ref{chi-X-Nf40}.

The transition we have identified for $\gamma = 1$ can be naturally
associated with the charged FP of the SU($N_c$) field theory
(\ref{cogau}). A conclusive proof would require a detailed analysis of
the gauge correlations. However, note that such a FP disappears as
$\gamma$ is decreased towards zero and is not present in the matrix
model in which the gauge fields are integrated out, confirming that
gauge fields do indeed play a role. It would be interesting to compare
the estimates of the critical exponents with the large-$N_f$
predictions computed in the gauge field theory---these results are not
available at present---as this would provide a more quantitative check
of the identification.

As a final check, we have studied the behavior of the model for
$\gamma = \infty$, to exclude that the observed behavior for $\gamma =
1$ is simply a crossover effect due to the presence of a continuous
transition in the infinite-$\gamma$ matrix model.  We recall that the
model in the $\gamma\to\infty$ limit becomes equivalent to the lattice
scalar model (\ref{hfixedlengthgammainf}), which can have continuous
transitions for sufficiently large $N_f$, and in particular for
$N_f=40$, see Sec.~\ref{particase}.  MC simulations of the ungauged
matrix model provide evidence of a phase transition for $\beta_c
\approx 1.00$. In Fig.~\ref{URxi-Nf40-gammainf} we report $U$ versus
$R_\xi$. We observe excellent scaling, indicating that the transition
is continuous. The curve we obtain is quite different from the one
obtained for $\gamma=1$, see Fig.~\ref{nc2nf40v1g1}. For instance, in
the matrix model the maximum of $U$ is approximately 1.007, which is
significantly smaller than the value obtained for $\gamma = 1$, see
the inset in the upper panel of Fig.~\ref{URxi-X-Nf40}.

We have also determined the exponents for the matrix model.  Although
we only have data for $8\le L \le 16$, scaling corrections are small.
Analyzing the data as we did for $\gamma=1$, we obtain
\begin{equation}
\beta_c = 1.0079(4)\,, \quad
\nu = 0.975(5)\,, \quad
\eta_Q = 1.147(5)\,.
\end{equation}
Note that $\eta_Q$ is the critical exponent associated with the
composite operator $Q$ and it should not be confused with the exponent
$\eta$ that characterizes the critical behavior of the correlations of
the fundamental field $\Phi^{af}$, that are well defined in the
ungauged model. The estimates of the exponents are very different from
those obtained for finite $\gamma$, again excluding that the results
for $\gamma=1$ are a crossover due to the presence of a continuous
transition in the ungauged matrix model.

\section{Conclusions}
\label{conclu}

We have investigated how nonabelian global and gauge symmetries shape
the phase diagram of 3D lattice gauge theories. We consider a model
with SU($N_c$) local invariance and U($N_f$) global invariance, in
which the scalar fields transform under the fundamental representation
of both groups. We use a standard formulation with nearest-neighbor
couplings \cite{Wilson-74}, considering the most general quartic
scalar potential compatible with the given global and gauge symmetry,
cf.~Eq.~(\ref{potential}).  We determine the low-temperature Higgs
phases and the nature of the phase transitions, as a function of the
parameter $v$ entering the quartic potential, defined in
Eq.~(\ref{hfixedlength}).  This study extends the one reported in
Refs.~\cite{BPV-19,BPV-20-su} for maximally symmetric scalar
potentials, corresponding to fixing $v=0$. We show that such an
extension to multiparameter quartic parameters give rise to various
notable scenarios, characterized by different low-temperature Higgs
phases.

The analysis of the minimum-energy configurations allows us to
determine the main features of the phase diagram. We determine the
ordered Higgs phases, their global and gauge symmetry-breaking pattern,
and the nature
of the transition lines between the various phases. These features
depend on the scalar-potential parameter $v$ and on the number of
colors and flavors, $N_c$ and $N_f$, respectively.  We observe
qualitative differences between the cases $N_c=2$ and $N_c>2$, and the
cases $N_f<N_c$, $N_f=N_c$ and $N_f>N_c$, as sketched in
Figs.~\ref{phdianf2nc2}, \ref{phdianfgtnc2}, \ref{phdianfltncge3},
\ref{phdianfeqncge3}, and \ref{phdianfgtncge3}.  In particular, for
$N_f \ge N_c$ the phase diagram presents two distinct Higgs phases,
associated with different global and gauge symmetry-breaking patterns.

To check the theoretical arguments, we performed numerical MC
simulations for $N_c = 2$.  For $v=0$ and any $N_f$, the model is
invariant under a larger symmetry group, namely Sp($N_f$)/${\mathbb
  Z}_2$.  Therefore, a first-order transition line is expected on the
$v=0$ axis for sufficiently large values of $\beta$, separating two
different low-temperature ordered phases corresponding to $v>0$ and
$v<0$, respectively, see Figs.~\ref{phdianf2nc2} and
\ref{phdianfgtnc2}. In particular, for $N_f=2$, the global symmetry of
the model with $v=0$ enlarges to $\hbox{Sp(2)} \simeq \hbox{O(5)}$,
leading to the emergence of an O(5) multicritical point, where the
continuous transition lines extending within the regions $v>0$ and
$v<0$ meet. According to the theoretical arguments reported in
Secs.~\ref{gausym} and \ref{phadia}, for $v>0$ the transition line
belongs to the XY universality class---the corresponding order
parameter is the determinant of the scalar fields, see
Eqs.~(\ref{Yphase}) and (\ref{Ydet})--- while, for $v < 0$, it belongs
to the O(3) universality class, being associated with the condensation
of the gauge-invariant bilinear operator defined in Eq.~(\ref{qdef}).
The FSS analyses of the numerical data support these theoretical
predictions, thus conferming the phase diagram sketched in
Fig.~\ref{phdianf2nc2}.

We also present results for larger values of $N_f$, focusing on the
phase behavior for $v>0$, whose nature is unknown, see
Fig.~\ref{phdianfgtnc2} and the corresponding discussion in
Sec.~\ref{phadia}. In particular, we address the question of the
existence of transitions that can be associated with the stable
charged FP that is present in the scalar SU(2) gauge field
theory---the corresponding Lagrangian is reported in
Eq.~(\ref{cogau})---for large values of $N_f$.

This issue has been recently addressed in the Abelian-Higgs field
theory characterized by a local U(1) and a global U($N_f$)
symmetry. Field theory predicts the existence of a stable charged FP
for a sufficiently large number of
components~\cite{HLM-74,DHMNP-81,FH-96,YKK-96,MZ-03,KS-08,IZMHS-19}.
In the $\epsilon$-expansion approach, such a FP only exists for $N_f >
N^*_f(d)$, where $d$ is the space dimension.  In $d=4$ dimensions,
$N^*_f(4) \approx 183$. However, corrections in the expansion in
powers of $\epsilon=4-d$ are large and four-loop results provide a
significantly smaller estimate in $d=3$, $N^*_f(3) \approx 12$. Recent
numerical work on the 3D lattice Abelian-Higgs model with noncompact
gauge fields \cite{BPV-21-nc} identified a transition line along which
critical exponents are in quantitative agreement with the field theory
large-$N_f$ predictions: these transitions can therefore be associated
with the charged FP.  These results provided the
estimate~\cite{BPV-21-nc} $N^*_f(3) = 7(2)$, confirming that the large
value in four dimensions, $N^*_f(4) \approx 183$, is quantitatively
not relevant for the 3D case.  It is worth noting that the charged FP
is also relevant for some transitions occuring in the compact
Abelian-Higgs model when the scalar matter has a charge larger than
one~\cite{BPV-20-hc}.

As it occurs in the scalar U(1) field theory, SU($2$) field theories
have a stable charged FP in the region $v>0$ for $N_f > N^*_f(d)$. Close to
four dimensions, $N^*(d)$ is very large, indeed
$N^*_f(4) \approx 376$, see
Sec.~\ref{sft}. However, it is conceivable that the critical value
$N^*_f(3)$ in three dimensions is significantly smaller than the
four-dimensional one, as it occurs in the Abelian-Higgs models.  To
check whether 3D SU(2) lattice models undergo transitions associated
with the field-theory charged FP, we have performed simulations for
two large number of components, $N_f=20$ and $N_f=40$.  For $N_f = 20$
we have only evidence of first-order transitions. A continuous
transition is instead observed for $N_f = 40$, $\gamma = 1$, and $v=1$.
The transition becomes of first order in the infinite-gauge coupling
limit ($\gamma \to 0$), in which gauge fields can be integrated out,
confirming that the gauge dynamics is relevant for the existence of
the continuous transition.  This leads us to conjecture that the continuous
transition observed for $N_f=40$ and finite $\gamma>0$
is associated with the charged FP of the SU($N_c$) field
theory with Lagrangian (\ref{cogau}). If the association is correct,
our results allow us to estimate $N^*_f$ in three dimensions. The
critical value $N^*_f(3)$ is large, $20<N^*_f(3)<40$, but still
significantly smaller that the four-dimensional value.

It is clear that significant additional work is needed to fully
clarify this issue. On the numerical side, a detailed analysis of
gauge correlations at the transition is clearly required, while on the
field-theory side it would be important to have quantitative
predictions for universal quantities, for instance, for the critical
exponents. Indeed, this would allow us to perform a more quantitative
comparison between the numerical results obtained in the simulation of
the lattice gauge model and the corresponding SU($N_c$) field theory
predictions.  A complete understanding of this issue is fundamental to
clarify if and how the nonabelian gauge field theory can be realized
in 3D statistical models sharing the same global and local symmetries.

\emph{Acknowledgement}.  Numerical simulations have been performed using
the CSN4 cluster of the Scientific Computing Center at INFN-PISA
and the Green Data Center of the University of Pisa.

\appendix

\section{Effective model for $N_c=2$} 
\label{AppA}

In this Appendix we briefly discuss the effective scalar model that
can be obtained for $\gamma = 0$ be integrating out the gauge fields.
We will use the results of Ref.~\cite{BRT-81} for SU($N$) link
integrals.  We define
\begin{equation}
S^{ab} = - {1\over2} N_f\beta \sum_f \Phi_{{\bm x} + \hat{\mu}}^{af}  
   (\Phi^{bf}_{\bm x})^* 
\end{equation}
and the invariant combination
\begin{eqnarray}
&& K_{{\bm x},\mu} = 
\hbox{Tr } S S^\dagger + \hbox{det} S + \hbox{det} S^\dagger 
\\
&& \quad = {1\over 4} N_f\beta^2 + 
  {1\over 4} N_f^2 \beta^2 \times \nonumber \\
&& \qquad \sum_{fg} 
   ( Q^{fg}_{{\bm x} + \hat{\mu}} Q^{fg}_{\bm x} + 
     {1\over2} \bar{Y}^{fg}_{{\bm x} + \hat{\mu}} Y^{fg}_{\bm x} + 
     {1\over2} Y^{fg}_{{\bm x} + \hat{\mu}} \bar{Y}^{fg}_{\bm x} ).
\nonumber 
\end{eqnarray}
Then, we obtain 
\begin{equation}
\int [dU] e^{-\beta H_K} = 
C 
\exp \sum_{{\bm x},\mu} \ln [I_1(2 K_{{\bm x},\mu})/\sqrt{K_{{\bm x},\mu}}],
\end{equation}
where $C$ is an irrelevant constant and $I_1(x)$ is a modified Bessel
function.  Since
\begin{equation}
\hbox{Tr } (\Phi^\dagger \Phi)^2 = \hbox{Tr } Q^2 - {1\over N_f} = 
   \hbox{Tr }\bar{Y} Y + 1 - {1\over N_f} , 
\end{equation}
we see that, for any $N_f$, in the absence of the gauge coupling, the
gauge model is equivalent to a matrix model for the order parameters
$Q$ and $Y$.  Note also that $K_{{\bm x},\mu}$ can be expressed in
terms of the Sp($N_f$) order parameter defined in
Ref.~\cite{BPV-20-su}, explicitly showing the larger symmetry of the
model for $v=0$.

We can also use these expressions to discuss the large-$\beta$ limit.
In this case we have translation invariance---the fields do not depend
on ${\bm x}$. If we use the singular value decomposition
(\ref{singdec}), we find that $K$ becomes independent of the scalar
fields, namely
\begin{equation}
K= {\beta^2 N_f^2\over 4} (w_1^2 + w_2^2)^2 = {\beta^2 N_f^2\over 4}.
\end{equation}
This result proves that for $v = 0$ and $\beta \to \infty$ the scalar
fields are uniformly distributed, as already shown in
Ref.~\cite{BPV-20-su}. Moreover, the scalar kinetic term is irrelevant
in determining the Higgs phases at low temperature: they are uniquely
fixed by the scalar potential.

\section{Monte Carlo simulations}
\label{mcsim}

We performed MC simulations on cubic lattices with periodic boundary
conditions. We used two different updates of the complex scalar field
$\Phi^{af}$. The first one is a standard Metropolis update
\cite{Metropolis:1953am} that rotates two randomly chosen elements of
$\Phi_{\bm x}^{ab}$ (denoted by $\phi_1$ and $\phi_2$ in the
following). More precisely the proposed update is
\begin{equation}
\begin{aligned}
\phi'_1 &= \cos\theta_1 e^{i\theta_2} \phi_1 + \sin\theta_1 e^{i\theta_3}\phi_2\\
\phi'_2 &= 
   -\sin\theta_1 e^{i\theta_2}\phi_1 + \cos\theta_1 e^{i\theta_3}\phi_2 \,,
\end{aligned}
\label{metro_rotation}
\end{equation}
where the angles $\theta_i$ are uniformly distributed in
$[-\alpha,\alpha]$, and the value of $\alpha$ are chosen to obtain an
acceptance of approximately 30\%.  In the second update we propose the
change
\begin{equation}
  \Phi'_{\bm x} = \frac{2 \mathrm{Re}\Tr(\Phi^\dagger_{\bm x} S_{\bm x})}
       {\Tr( S^\dagger_{\bm x} S_{\bm x})}S_{\bm x} 
- \Phi_{\bm x}\,,
\label{generalized-overrelaxed}
\end{equation}
where $S_{\bm x}$ is the matrix
\begin{equation}
    S_{\bm x} = \sum_\mu 
    (U_{{\bm x}, \mu} \Phi_{{\bm x}+\hat{\mu}} +
    U^\dagger_{{\bm x}-\hat{\mu}, \mu} \Phi_{{\bm x}-\hat{\mu}}) \,.
\end{equation}
Such a deterministic update satisfies detailed balance (since it is
involutive), and for $v=0$ would be an overrelaxation step
\cite{Creutz:1987xi}. For $v\neq 0$ this move is accepted or rejected
using a standard Metropolis test, and for the parameters used in this
work a typical value of the corresponding acceptance rate is 90\%.
Link variables were updated using the Metropolis algorithm, with the
proposed update $U_{{\bm x},\mu}\to V U_{{\bm x},\mu}$, where $V$ is
an SU($N_c$) matrix close to the identity and $V$ or $V^{\dag}$ were
used with a 50\% probability to ensure detailed balance.  Also in this
case the maximal distance of $V$ from the identity matrix was chosen
in such a way to have an average 30\% acceptance ratio.

We call lattice iteration a series of 10 lattice sweeps in which we
sequentially update the scalar field on all the sites and the gauge
field on all the links. In 9 lattice sweeps we use the
pseudo-overrelaxed update with proposal
(\ref{generalized-overrelaxed}), while in 1 sweep we use the update
based on the proposal (\ref{metro_rotation}). This ratio of 1 to 9 was
kept fixed for all the cases studied in this work, since we verified
the autocorrelation times to be small enough for our purposes, and we
did not pursued any further parameter optimization.

Measures where performed after every lattice iteration, and for the
largest lattice sizes typical statistics of our runs were $\approx 3
\times 10^6$ measures in the case of two-flavor models, and $\approx 8
\times 10^5$ and $\approx 4 \times 10^5$ for $N_f=20$ and $N_f=40$
respectively.  To analyze data and estimate error bars we used
standard blocking and jackknife techniques, and the maximum blocking
size adopted was of the order of $10^{3}$ data.


\begin{thebibliography}{99}

\bibitem{Wilson-74} K. G. Wilson, Confinement of quarks, Phys. Rev. D
  {\bf 10}, 2445 (1974).

\bibitem{ZJ-book} J. Zinn-Justin, 
  {\em Quantum Field Theory and Critical Phenomena}, 
  fourth edition (Clarendon Press, Oxford, 2002).

\bibitem{Weinberg-book} S. Weinberg, {\em The Quantum Theory of
  Fields}, (Cambridge University Press, 2005).

\bibitem{Wegner-71} F. J. Wegner. Duality in generalized Ising models
  and phase transitions without local order parameters J. Math.
  Phys. {\bf 12}, 2259 (1971).

\bibitem{Anderson-book} P.~W.~Anderson, {\em Basic Notions of
  Condensed Matter Physics}, (The Benjamin/Cummings Publishing
  Company, Menlo Park, California, 1984).
  
\bibitem{Sachdev-19} S. Sachdev, Topological order, emergent gauge
  fields, and Fermi surface reconstruction, Rep. Prog. Phys. {\bf 82},
  014001 (2019).

\bibitem{GG-72} H. Georgi and S. L. Glashow, Unified weak and
  electromagnetic interactions without neutral currents,
  Phys. Rev. Lett. {\bf 28}, 1494 (1972).

\bibitem{OS-78} K.~Osterwalder and E.~Seiler, Gauge Field Theories on
  the Lattice, Ann. Phys. (NY) {\bf 110}, 440 (1978).

\bibitem{FS-79} E. Fradkin and S. Shenker, Phase diagrams of lattice
  gauge theories with Higgs fields, Phys. Rev. D {\bf 19}, 3682
  (1979).

\bibitem{DRS-80} S.~Dimopoulos, S.~Raby, and L.~Susskind, Light
  Composite Fermions, Nucl. Phys. B {\bf 173}, 208 (1980).

\bibitem{BN-87} C. Borgs and F. Nill, The Phase Diagram of the Abelian
  Lattice Higgs Model. A Review of Rigorous Results,
  J. Stat. Phys. {\bf 47}, 877 (1987).

\bibitem{BPV-19} C. Bonati, A. Pelissetto, and E. Vicari, Phase
  Diagram, Symmetry Breaking, and Critical Behavior of
  Three-Dimensional Lattice Multiflavor Scalar Chromodynamics,
  Phys. Rev. Lett. {\bf 123}, 232002 (2019).

\bibitem{BPV-20-su} C. Bonati, A. Pelissetto, and E. Vicari,
  Three-dimensional lattice multiflavor scalar chromodynamics:
  Interplay between global and gauge symmetries, Phys. Rev. D {\bf
    101}, 034505 (2020).

\bibitem{BPV-20-on} C. Bonati, A. Pelissetto, and E. Vicari,
  Three-dimensional phase transitions in multiflavor scalar SO($N_c$)
  gauge theories, Phys. Rev. E {\bf 101}, 062105 (2020).
  
\bibitem{SSST-19} S. Sachdev, H. D. Scammell, M. S. Scheurer, and
  G. Tarnopolsky, Gauge theory for the cuprates near optimal doping,
  Phys. Rev. B {\bf 99}, 054516 (2019).

\bibitem{SPSS-20} H. D. Scammell, K. Patekar, M. S. Scheurer, and
  S. Sachdev, Phases of SU(2) gauge theory with multiple adjoint Higgs
  fields in 2$+$1 dimensions, Phys. Rev. B {\bf 101}, 205124 (2020).

\bibitem{BFPV-21-3d} C. Bonati, A. Franchi, A. Pelissetto, and
  E. Vicari, Three-dimensional lattice SU($N_c$) gauge theories with
  multiflavor scalar fields in the adjoint representation, Phys. Rev B
  {\bf 114}, 115166 (2021).

\bibitem{Nadkarni:1989na} 
  S.~Nadkarni,
  The SU(2) Adjoint Higgs Model in Three dimensions,
  Nucl. Phys. B {\bf 334}, 559 (1990).

\bibitem{Kajantie:1993ag} 
  K.~Kajantie, K.~Rummukainen and M.~E.~Shaposhnikov,
  A Lattice Monte Carlo study of the hot electroweak phase transition,
  Nucl. Phys. B {\bf 407}, 356 (1993).

\bibitem{Buchmuller:1994qy} 
  W.~Buchm\"uller and O.~Philipsen,
  Phase structure and phase transition of the SU(2) Higgs model 
  in three-dimensions,
  Nucl. Phys. B {\bf 443}, 47 (1995).
 
\bibitem{Kajantie:1996mn} 
  K.~Kajantie, M.~Laine, K.~Rummukainen, and M.~E.~Shaposhnikov,
  Is there a hot electroweak phase transition at $m_H \gtrsim m_W$?,
  Phys. Rev. Lett.  {\bf 77}, 2887 (1996).
  

\bibitem{Hart:1996ac} 
  A.~Hart, O.~Philipsen, J.~D.~Stack, and M.~Teper,
  On the phase diagram of the SU(2) adjoint Higgs model in (2+1)-dimensions,
  Phys. Lett. B {\bf 396}, 217 (1997).

\bibitem{BPV-21-nc} C. Bonati, A. Pelissetto, and E. Vicari, Lattice
  Abelian-Higgs models with noncompact gauge field, Phys. Rev. B {\bf
    103}, 085104 (2021).

\bibitem{BPV-20-hc} C. Bonati, A. Pelissetto, and E. Vicari, Higher-charge
  three-dimensional compact lattice Abelian-Higgs models, Phys. Rev. E
  {\bf 102}, 062151 (2020).
  
\bibitem{PV-02} A. Pelissetto and E. Vicari, Critical phenomena and
  renormalization group theory, Phys. Rep. {\bf 368}, 549 (2002).

\bibitem{CLB-86}
M. S. S. Challa, D. P. Landau, and K. Binder,
Finite-size effects at temperature-driven first-order transitions
Phys. Rev. B {\bf 34}, 1841 (1986).

\bibitem{VRSB-93}
K. Vollmayr, J. D. Reger, M. Scheucher, and K. Binder,
Finite size effects at thermally-driven first order phase transitions: A
phenomenological theory of the order parameter distribution
Z.~Phys.~B {\bf 91} 113 (1993).  

\bibitem{CPPV-04} P. Calabrese, P. Parruccini, A. Pelissetto,
  and E. Vicari, Critical behavior of O(2)$\otimes$O($N$)-symmetric
  models, Phys. Rev. B {\bf 70}, 174439 (2004).

\bibitem{PV-19} A. Pelissetto and E. Vicari, Three-dimensional
  ferromagnetic CP$^{N-1}$ models, Phys. Rev. E {\bf 100}, 022122
  (2019).

\bibitem{PV-19-AH3d} A. Pelissetto and E. Vicari, 
Multicomponent compact Abelian-Higgs lattice models, 
Phys. Rev. E {\bf 100}, 042134 (2019).

\bibitem{PV-20-ln} A. Pelissetto and E. Vicari, 
Large-$N$ behavior of three-dimensional lattice CP$^{N-1}$ models, 
J. Stat. Mech. (2020) 033209.

\bibitem{NCSOS-11}
A. Nahum, J. T. Chalker, P. Serna, M. Ortu\~no, and A. M. Somoza,
3D Loop Models and the CP$^{N-1}$ Sigma Model,
Phys.  Rev.  Lett. {\bf 107}, 110601 (2011).

\bibitem{NCSOS-13}
A. Nahum, J. T. Chalker, P. Serna, M. Ortu\~no, and A. M. Somoza,
Phase transitions in three-dimensional loop models and the CP$^{N-1}$ 
sigma model, Phys. Rev. B {\bf 88}, 134411 (2013).

\bibitem{KS-12}  
R. K. Kaul and A. W. Sandvik,
Lattice Model for the SU($N$) Ne\`el to valence-bond solid quantum phase
transition at large $N$,
Phys. Rev. Lett. {\bf 108},  137201 (2012) 

\bibitem{IZMHS-19} 
B. Ihrig, N. Zerf, P. Marquard, I. F. Herbut, and M. M. Scherer, 
Abelian Higgs model at four loops, fixed-point
collision and deconfined criticality, Phys. Rev. B {\bf 100}, 134507 (2019).

\bibitem{MZ-03} M. Moshe and J. Zinn-Justin, Quantum field theory in
  the large $N$ limit: A review, Phys. Rep. {\bf 385}, 69 (2003).

\bibitem{HLM-74} 
B. I. Halperin, T. C. Lubensky, and S. K. Ma,
  First-Order Phase Transitions in Superconductors and Smectic-A
  Liquid Crystals, Phys. Rev. Lett. {\bf 32}, 292 (1974).

\bibitem{DHMNP-81} P. Di Vecchia, A. Holtkamp, R. Musto, F. Nicodemi,
  and R. Pettorino, Lattice CP$^{N-1}$ models and their large-$N$
  behaviour, Nucl. Phys. B {\bf 190}, 719 (1981).

\bibitem{FH-96} 
R. Folk and Y. Holovatch, On the critical fluctuations
  in superconductors, J. Phys. A {\bf 29}, 3409 (1996).

\bibitem{YKK-96} V. Yu. Irkhin, A. A. Katanin, and M. I. Katsnelson,
  $1/N$ expansion for critical exponents of magnetic phase transitions
  in the $CP^{N-1}$ model for $2<d<4$, Phys. Rev. B {\bf 54}, 11953
  (1996).

\bibitem{KS-08} R. K. Kaul and S. Sachdev, Quantum criticality of U(1)
  gauge theories with fermionic and bosonic matter in two spatial
  dimensions, Phys. Rev. B {\bf 77}, 155105 (2008).

\bibitem{CP-inprep} R. Cipolloni, Sapienza Master thesis (2022);
  S. Rulli, Master thesis, University of Pisa (2022).


\bibitem{PW-84} R. D. Pisarski and F. Wilczek, Remarks on the chiral
  phase transition in chromodynamics, Phys. Rev. D {\bf 29}, 338
  (1984).

\bibitem{BPV-03} A. Butti, A. Pelissetto, and E. Vicari, On the nature
  of the finite-temperature transition in QCD, J. High Energy
  Phys. {\bf 08}, 029 (2003).

\bibitem{PV-13} A. Pelissetto and E. Vicari, Relevance of the axial
  anomaly at the finite-temperature chiral transition in QCD,
  Phys. Rev. D {\bf 88}, 105018 (2013).

\bibitem{CP-04} P. Calabrese and P. Parruccini, Five-loop epsilon
  expansion for U($n$)$\otimes$U($m$) models: finite-temperature phase
  transition in light QCD, J. High Energy Phys. 05 (2004) 018.

\bibitem{Georgi-book} H.~Georgi, {\em Weak interactions and modern particle
  theory}, (The Benjamin/Cummings Publishing Company, Menlo Park, California,
  1984).

\bibitem{AY-94} P. Arnold and L. G. Yaffe, The $\epsilon$ expansion
  and the electroweak phase transition, Phys. Rev. D {\bf 49}, 3003
  (1994).

\bibitem{DP-14} 
P.~S.~Bhupal Dev and A.~Pilaftsis, Maximally Symmetric
  Two Higgs Doublet Model with Natural Standard Model Alignment, JHEP
  {\bf 1412}, 024 (2014); (Erratum) JHEP {\bf 1511}, 147 (2015).

\bibitem{WNMXS-17} C. Wang, A. Nahum, M. A. Metlitski, C. Xu, and
  T. Senthil, Deconfined Quantum Critical Points: Symmetries and
  Dualities, Phys. Rev. X {\bf 7}, 031051 (2017).

\bibitem{CPV-03} P. Calabrese, A. Pelissetto, and E. Vicari,
  Multicritical behavior of ${\rm O}(n_1)\oplus {\rm
    O}(n_2)$-symmetric systems, Phys. Rev. B {\bf 67}, 054505 (2003).

\bibitem{HPV-05} M. Hasenbusch, A. Pelissetto, and E. Vicari,
  Instability of the O(5) critical behavior in the SO(5) theory of
  high-$T_c$ superconductors, Phys. Rev. B {\bf 72} 014532 (2005).



\bibitem{CHPV-06} M. Campostrini, M. Hasenbusch, A. Pelissetto, and
  E. Vicari, Theoretical estimates of the critical exponents of the
  superfluid transition in $^4$He by lattice methods, Phys. Rev. B {\bf 74},
  144506 (2006).

\bibitem{Hasenbusch-19} M. Hasenbusch, Monte Carlo study of an
  improved clock model in three dimensions, Phys. Rev. B {\bf 100},
  224517 (2019).  

\bibitem{CLLPSSV-20} S. M. Chester, W. Landry, J. Liu, D. Poland,
  D. Simmons-Duffin, N. Su, and A. Vichi, Carving out OPE space and
  precise O(2) model critical exponents,
J. High Energy Phys. {\bf 06}, 142 (2020).

\bibitem{BPV-21-coAH} C.~Bonati, A.~Pelissetto and E.~Vicari, Lattice
  gauge theories in the presence of a linear gauge-symmetry breaking,
  Phys. Rev. E {\bf 104}, 014140 (2021).

\bibitem{Hasenbusch-20} M. Hasenbusch, Monte Carlo study of a
  generalized icosahedral model on the simple cubic lattice,
  Phys. Rev. B {\bf 102}, 024406 (2020).

\bibitem{Chester-etal-20-o3} S. M. Chester, W. Landry, J. Liu,
D. Poland, D. Simmons-Duffin, N. Su, and A. Vichi,
Bootstrapping Heisenberg magnets and their cubic instability,
  arXiv:2011.14647.

\bibitem{KP-17}
M. V. Kompaniets and E. Panzer, Minimally subtracted six-loop 
renormalization of $\phi^4$-symmetric theory and critical exponents, 
Phys. Rev. D {\bf 96}, 036016 (2017).

\bibitem{HV-11} M. Hasenbusch and E. Vicari, Anisotropic perturbations
  in 3D O(N) vector models, Phys. Rev. B {\bf 84}, 125136 (2011).

\bibitem{CHPRV-02} M. Campostrini, M. Hasenbusch, A. Pelissetto,
  P. Rossi, and E. Vicari, Critical exponents and equation of state of
  the three-dimensional Heisenberg universality class, Phys. Rev. B
  {\bf 65}, 144520 (2002).

\bibitem{GZ-98} R. Guida and J. Zinn-Justin, Critical exponents of the
  $N$-vector model, J. Phys. A {\bf 31}, 8103 (1998).

\bibitem{BRT-81}
R. Brower, P. Rossi, and C.-I.Tan,
The external field problem for QCD,
Nucl. Phys. B {\bf 190}, 699 (1981).

\bibitem{Metropolis:1953am} N.~Metropolis, A.~W.~Rosenbluth,
  M.~N.~Rosenbluth, A.~H.~Teller, and E.~Teller, Equation of state
  calculations by fast computing machines, J.\ Chem.\ Phys.\ {\bf 21},
  1087 (1953).  

\bibitem{Creutz:1987xi} M.~Creutz, Overrelaxation and Monte Carlo
  Simulation, Phys.\ Rev.\ D {\bf 36}, 515 (1987).

\end{thebibliography}
\end{document}